\documentclass[aps,onecolumn,superscriptaddress,floatfix,pt12,color]{revtex4}

\usepackage{graphicx}
\usepackage{epsfig}
\usepackage{amssymb}

\usepackage[dvips]{color}
\definecolor{Red}{rgb}{1.00, 0.00, 0.00}

\begin{document}
\newcommand{\be}{\begin{eqnarray}}
\newcommand{\ee}{\end{eqnarray}}
\newcommand\del{\partial}
\newcommand\nn{\nonumber}
\newcommand{\Tr}{{\rm Tr}}
\newcommand{\mat}{\left ( \begin{array}{cc}}
\newcommand{\emat}{\end{array} \right )}
\newcommand{\vect}{\left ( \begin{array}{c}}
\newcommand{\evect}{\end{array} \right )}
\newcommand{\tr}{\rm Tr}
\def\conj#1{{{#1}^{*}}}
\newcommand\hatmu{\hat{\mu}}
\newcommand\noi{\noindent}

\voffset 2cm

\title{Distributions of the Phase Angle of the Fermion Determinant in QCD}

\author{M.P. Lombardo}
\affiliation{INFN-Laboratori Nazionali de Frascati, I-00044, Frascati (RM), Italy} 
\author{K. Splittorff}
\affiliation{The Niels Bohr Institute, Blegdamsvej 17, DK-2100, Copenhagen
  {\O}, Denmark} 
\author{J.J.M. Verbaarschot}
\affiliation{Department of Physics and Astronomy, SUNY, Stony Brook,
 New York 11794, USA}

\date   {\today}
\begin  {abstract}

The distribution of the phase angle and the magnitude of the fermion 
determinant as well as its correlations with the baryon number and 
the chiral condensate are studied for
QCD at non zero quark chemical potential. Results are derived to one-loop
order in chiral perturbation theory.
We find that the
distribution of the phase angle is Gaussian for small chemical potential and 
a periodic Lorentzian when the quark mass is 
inside the support of the Dirac spectrum.
The baryon number and chiral condensate are computed as a function of the
phase of the fermion determinant and we discuss the severe cancellations
which occur upon integration over the angle. 
We compute the
distribution of the magnitude of the fermion determinant
as well as the baryon number and chiral
condensate at fixed  magnitude. 

Finally, we consider QCD in one Euclidean dimension where it is shown
analytically, starting from the fundamental QCD partition function, that the 
distribution of the phase of the fermion determinant is a periodic 
Lorentzian when the quark mass is inside the 
spectral density of the Dirac operator.
\end{abstract}

\maketitle
\newpage
 
\section{Introduction}

The phase diagram of strongly interacting matter is expected to show
several phases as a function of the temperature and the baryon
chemical potential. Matter in  nuclei, in compact stars and
in the early universe are in  different parts of the phase diagram 
and
large experimental and theoretical efforts have been
invested to understand their properties.
Of particular intense interest is
the critical  end-point.  Its existence is expected mainly on the findings 
of model studies that the baryon density is discontinuous as a function of
the chemical potential 
\cite{endpoint}. Lattice QCD, which has allowed us to determine the nature of
the phase transition at zero baryon chemical potential \cite{Nature}, appears 
to be the natural tool to study the non-perturbative phenomena which take
place near the endpoint. However, probabilistic lattice QCD methods 
are  not directly applicable at
nonzero baryon chemical potential: Monte Carlo 
importance sampling, which is at the core of Lattice QCD computations,
requires that the
Euclidean action is real.  At non zero chemical potential, though, 
the quark 
determinant is complex. This severe obstacle is known as 
{\sl the sign problem}.

Recent numerical progress in understanding 
the phase diagram of strongly interacting
matter at nonzero chemical potential has reopened the field. Not only has it
been understood that the location of the endpoint in the $(\mu,T)$-plane
is extremely sensitive to the quark mass \cite{deFPh}, it may also be that  
the dependence of the endpoint on quark mass is very different from what was
commonly accepted \cite{deFPh}. Because of the sign problem these
conclusions where reached from analytic continuations of lattice simulations
carried out at imaginary values of the chemical potential. Such an
extrapolation \cite{owe1,maria,owe2,ERL} is not without pitfalls.     
It has recently been demonstrated \cite{BJ} that utmost care 
should be taken when attempting to extract information on the critical
endpoint from a Taylor expansion at $\mu=0$
\cite{gupta,Allton1,Allton2,Allton3,Endrodi:2009sd}. Moreover, it was demonstrated
in \cite{deFSW} that the numerical implementation of the
re-weighting approach \cite{Glasgow,fodor1,fodor2} is extremely delicate even
at small values of the chemical potential.

Lately alternative numerical methods such as the density of states
method and the complex Langevin method have been explored. 
Despite early reports of its failure \cite{KW,FOC,AFP},
the complex Langevin method has been shown to be able to deal with sign
problems in simple models and for a gas of relativistic bosons \cite{aarts}. 
On the analytical front, the severity
of the sign problem was analyzed for QCD at low energy and for models
of the QCD partition function 
\cite{exp2ith-letter,phase-long,1loop,Han,BW}. The intricate connections
between  the sign
problem, chiral symmetry, and the Dirac spectrum, have been understood in the
$\epsilon$-regime of QCD \cite{OSV}.   

In the present work we focus on the density of states method
\cite{Gocksch,Azcoiti,AN,AANV,Schmidt,Ejiri}.   
In this approach one evaluates an observable numerically for a fixed
given quantity and thereby obtain the distribution of this observable
over the fixed quantity. The full expectation value of the observable
is then obtained by integration over the fixed quantity. 
This method has had some success when the baryon number,
the average plaquette  or the phase of the fermion determinant is kept 
fixed. In this paper we are particularly interested in the last approach
since it goes back to the root of the sign problem.
If we would know the exact distribution function of the phase of the fermion
determinant as well as  its correlations with physical observables,
the sign problem would have been solved: the delicate cancellations due to
the fluctuations of the phase could be realized exactly by an analytical
integration over the phase according to the distribution function
and its correlations.

We will use chiral perturbation theory to compute the distribution of the
phase of the fermion determinant
\be\label{rhoTh1} 
\langle\delta(\theta-\theta')\rangle_{N_f}d\theta
=\frac{\int dA |\det(D+\mu\gamma_0+m)|^{N_f} e^{iN_f\theta'}
  \delta(\theta-\theta') e^{-S_{\rm YM}}}
{\int dA |\det(D+\mu\gamma_0+m)|^{N_f} e^{iN_f\theta'} e^{-S_{\rm YM}}} d\theta .
\ee
Here $\theta'$ refers to the phase of the
fermion determinant.  It is a function of 
the gauge field configuration which we average over, i.e.
$\exp(2i\theta')=\det(D(A)+\mu\gamma_0+m)/{\det}^*(D(A)+\mu\gamma_0+m)$. 
Due to the sign problem the distribution of the phase is not real and
positive. The complex nature, however, is of the simplest possible form:
Since   
\be\label{th-distBasic}
 \langle \delta(\theta-\theta')\rangle_{N_f} = e^{i\theta N_f}\frac{Z_{|N_f|}}{Z_{N_f}}\langle \delta(\theta-\theta')\rangle_{|N_f|}.  
\ee
the $\theta$-distribution factorizes into  $\exp(i \theta N_f)$ and a real
and positive distribution.
Here, $Z_{N_f} $  is the $N_f$ flavor partition function and $Z_{|N_f|} $ is
the phase quenched $N_f$ flavor partition function. The subscripts $N_f$ 
and $ |N_f|$ refer to averages with respect to these two partition functions,
in this order. For $N_f = 2$ this relation reads
\be\label{th-distBasic}
 \langle \delta(\theta-\theta')\rangle_{1+1} = e^{2i\theta}\frac{Z_{1+1^*}}
{Z_{1+1}}\langle \delta(\theta-\theta')\rangle_{1+1^*},  
\ee
where here and below the subscript $1+1$ refers to
QCD with two ordinary flavors whereas the subscript $1+1^*$ refers to QCD
with one ordinary flavor and one conjugate flavor. By definition the fermion
determinant of a quark and a conjugate quark are each others complex 
conjugates  so that the 
total measure is real and positive. The $\theta$-distribution of the {\sl
  phase quenched} theory, $\langle \delta(\theta-\theta')\rangle_{1+1^*}$, is
necessarily real and positive. Moreover it is normalized to one. Also the
$\theta$-distribution of the full theory  $\langle
\delta(\theta-\theta')\rangle_{1+1}$ is normalized to one. On the r.h.s. of
(\ref{th-distBasic}), however, the ratio $Z_{1+1^*}/Z_{1+1}$ grows
exponentially fast with the volume so that the phase factor, $e^{2i\theta}$,
must lead to exponentially large cancellations.

\begin{figure}[t!]
  \unitlength1.0cm
  \epsfig{file=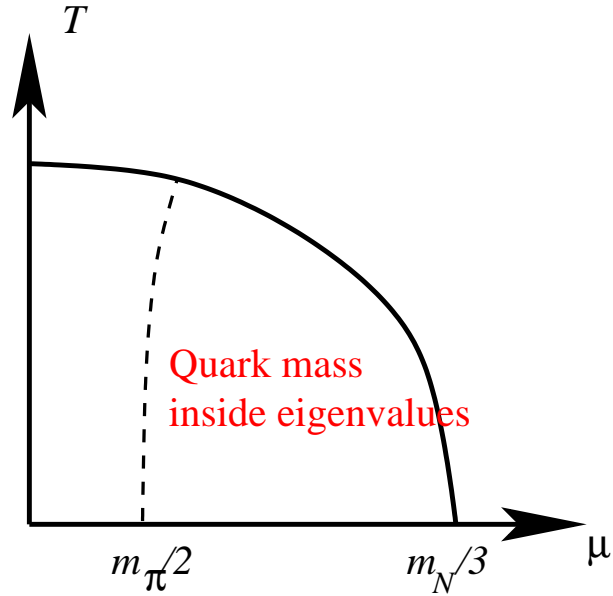,clip=,width=8cm}
  \caption{\label{fig:PD} A schematic picture of the phase diagram of QCD as
    a function of the quark chemical potential $\mu$ and the temperature $T$.
    Chiral symmetry is spontaneously broken below the full curve. The dashed
    curve indicates where the quark mass enters the Dirac spectrum. As this
    happens the nature of
    the sign problem changes. To the left of the dashed curve the distribution
    of phase of the fermion determinant is a periodic superposition of 
Gaussians whereas it is a periodic superposition of
    Lorentzians to the right of the dashed curve. We stress that the
    dashed curve does {\sl not} indicate a phase transition in QCD.}
\end{figure}

As we shall see it is essential to discuss separately the case when
$2\mu/m_\pi$ is small and the case when the quark
mass is inside the spectral density of the Dirac operator. 
We will show below that
the real and positive part of the $\theta$-distribution becomes a periodic
superposition of Gaussians
when the quark mass $m$ is outside the support of the Dirac spectrum. When
the quark mass is inside the support of the Dirac spectrum the sign problem
becomes much more severe \cite{OSV,1loop}. Figure \ref{fig:PD} gives a
schematic picture of the phase diagram of QCD as well as the region where the
quark mass is inside the spectral support of the Dirac operator. 
As we will show below, the $\theta$-distribution in   
this region is not only very wide/flat, it also changes shape into a periodic
superposition of  
Lorentzians. A hint of this dramatic change is already present in
(\ref{th-distBasic}). When the quark mass enters the spectral support of the
Dirac operator a phase transition occurs in the phase quenched theory while 
the full theory remains 
unaltered \cite{OSV}. The exponential growth of the ratio
$Z_{1+1^*}/Z_{1+1}$ with the volume is thus particularly rapid when the quark
mass enters the spectral support of the Dirac operator. (The link between the
Dirac spectrum of the full theory and the phases of the phase quenched theory
are discussed in detail in \cite{misha,SplitVerb2,AOSV,SpecPhase}).

The Gaussian shape of the $\theta$-distribution for small $\mu$ was first
observed numerically by Ejiri in \cite{Ejiri} where it is also argued that
this form is a natural consequence of the central limit theorem. The change
from the Gaussian to the Lorentzian form for larger values of $\mu$ therefore
suggests a breakdown of the conditions for the application
of the central limit theorem. To
cast further light on this we also compute the distribution of the phase, 
$\langle\delta(\theta-\theta')\rangle$, for lattice QCD in one Euclidean
dimension. As we will show, it is possible to derive the Lorentzian form of the
$\theta$-distribution directly from the one dimensional lattice QCD partition
function, when the quark mass is inside the support of the Dirac
spectrum.

\vspace{1mm}

In addition to the distribution of the phase of the fermion determinant we
also consider the direct dependence of observables $\cal{O}$ on the phase
$\theta$ through the distribution-function  
\be
 \langle \cal{O} \ \delta(\theta-\theta')\rangle.  
\ee
The integral over $\theta$ obviously gives the full expectation value
 $\langle {\cal
  O}\rangle$. The $\theta$-dependence of the observable shows if severe
cancellations take place in this integral. Furthermore, the distribution of the
observable with the phase allows us to address which range of the phase is
essential for the full expectation value of $\cal{O}$. 

We will compute the distribution of the baryon number
operator, its square as well as the distribution of the chiral
condensate over $\theta$. 
It is found that the distributions, $\langle \cal{O} \
\delta(\theta-\theta')\rangle$, take complex values and that drastic
cancellations occur when integrating over $\theta$.

\vspace{5mm}

This paper is organized as follows.
In section \ref{sec:phfactr1loop} and \ref{sec:thdistmu<0p5mpi} we
briefly recall a few facts about chiral perturbation theory which are
relevant for the calculation of the  average phase factor and the distribution
of the phase angle. 
Then we turn to the distributions of the baryon number (section
\ref{sec:baryon}), the off-diagonal susceptibility (section
\ref{sec:baryonSQ}) and the chiral condensate (section \ref{sec:Sigma}) over
the phase angle of the fermion determinant. 
These one-loop results are all valid for
$\mu<m_\pi/2$. Next we discuss the
distribution of the phase for an ensemble generated at $\mu=0$.  
The difference  in the
phase distribution for  $\epsilon$-counting rather than the $p$-counting
in pointed out in section \ref{sec:epsilon}. 
In section \ref{sec:Lorentzian} it is shown that the leading order
prediction for the $\theta$-distribution takes a Lorentzian shape for
$\mu>m_\pi/2$. The Lorentzian form is then obtained as an exact result for 
lattice QCD in one Euclidean dimension in section \ref{sec:1dQCD}. 
The remainder of
the paper discusses the radial distribution of the fermion determinant.

\section{1-loop chiral perturbation theory and the average phase factor}
\label{sec:phfactr1loop}

The first step towards obtaining the distribution of the phase is to
understand the average of the phase factor. In this section we review 
the calculation of the average
phase factor in chiral perturbation theory.

Chiral perturbation theory \cite{CPT} is the low energy effective theory of QCD in the
phase where chiral symmetry is broken spontaneously. It describes the
dynamics of the Goldstone modes, i.e. the pions and the kaons.    
We shall work in the so called $p$-expansion of chiral perturbation theory
where the small expansion parameter is  
\be
p \sim m_\pi \sim \mu \sim T \sim \frac{1}{L}.
\label{p-counting}
\ee

For $\mu<m_\pi/2$ the chemical potential modifies the pion propagator in the
standard way for relativistic bosons. 
The one-loop contribution to the free energy from a pair of charge conjugate
pions (the chemical potentials are therefore $\mu$ and $-\mu$) is thus
given by
\be
G_0(\mu,-\mu) \equiv - \sum_{p_{k\, \alpha}}
\log(|\vec p^2_{k\,\alpha} +m^2_\pi +(p_{k\, 0}-2i\mu)^2|^2) ,
\ee
where
\be
p_{k\,\alpha} = \frac {2\pi k_\alpha}{L_\alpha}, 
\qquad k_\alpha \quad {\rm integer}.
\ee
After a Poisson resummation this can be expressed as \cite{phase-long} 
\be
G_0(\mu,-\mu)
=-V \sum_{l_\alpha}\int \frac{d^d p}{(2\pi)^d } e^{iL_\alpha  p_\alpha l_\alpha} 
\log(|\vec p^2 +m^2_\pi +(p_{0}-2i \mu)^2|^2) ,
\ee
where the sum is over all integers. The thermodynamic limit is given
by the term $l_\alpha = 0$. Here, two facts about this term, which we denote 
by $G_0|_{V=\infty}$, are essential: {\sl i)} it is independent of $\mu$
 {\sl ii)} it includes the entire 1-loop divergence (see \cite{STV} for a
 discussion). In dimensional regularization it is given by
\be
G_0|_{V=\infty} =
\frac{2}{(4\pi)^{d/2}}\Gamma\left(-\frac{d}{2}\right)m_\pi^d .
\ee
The finite part of the 1-loop free energy, denoted by $g_0(\mu)$, contains
the sum over the terms with $l_\alpha \neq 0$. This results in the 
decomposition
\be
G_0(\mu,-\mu) = G_0|_{V=\infty} + g_0(\mu,-\mu).
\ee
If we wish to keep track of the 
leading $1/V$ corrections to the infinite volume result we have to evaluate
the sum over all
four components of the momentum. The finite, $\mu$, $L$ and $T$ dependent part
then reads \cite{phase-long}  (this expression generalizes the result of
\cite{HL} for $\mu = 0$ to nonzero chemical potential) 
\be
g_0(\mu,-\mu)
&=&
2 \int_0^\infty \frac{d\lambda}{\lambda^3}
e^{-m^2_\pi L^2\lambda /4\pi }
(\prod_{\alpha=0}^3 {\sum_{l_\alpha}} e^{-2\mu l_0 L_0\delta_{\alpha 0}} 
e^{ -\pi \frac{l_\alpha^2 L_\alpha^2}{\lambda L^2}} -1),
\label{g0finiteL}
\ee
where $l_\alpha$ runs over all integers and $L \equiv (L_0L_i^3)^{1/4}$.

When the  length of the box is considerably larger than the Compton
wavelength of the pion the sum over
momenta can be replaced by an integral, and the 1-loop
contribution to the free energy simplifies to the familiar expression
\be
g_0(\mu,-\mu)
& = & \frac{Vm_\pi^2T^2}{\pi^2}\sum_{n=1}^\infty 
\frac{K_2(\frac{m_\pi n}{T})}{n^2}\cosh(\frac{2\mu n}{T}).   
\label{g0familiar}
\ee 

As the simplest relevant example let us now consider the average phase factor
for the phase quenched theory. By definition we have that
\be
\langle e^{2i\theta'}\rangle_{1+1^*} = \frac{Z_{1+1}}{Z_{1+1^*}}. 
\ee
The phase quenched theory in the denominator is identical to QCD at nonzero  
chemical potential for the third component of isospin \cite{AKW}. This has an
immediate consequence: 
since the pions carry isospin charge but no baryon
charge the free energy of $Z_{1+1^*}$ depends on $\mu$ while the usual free 
energy of
$Z_{1+1}$ is independent of $\mu$ when evaluated in chiral
perturbation theory.
It is this dependence on the chemical potential which makes it possible to
compute the average phase factor in chiral perturbation theory despite the
fact that pions have baryon charge zero.

For small $\mu$ the leading (mean field) term in the chiral Lagrangian,
$2m\langle\bar\psi\psi\rangle V$, is identical in the two cases and hence the
phase factor is determined to leading order by the one-loop effect 
\be
\langle e^{2i\theta'}\rangle_{1+1^*} =
\frac{e^{G_0(\mu,\mu)}}{e^{G_0(\mu,-\mu)}} 
= e^{g_0(\mu=0)-g_0(\mu)}. 
\ee
With $p$-counting (\ref{p-counting}) we have that 
$g_0(\mu)-g_0(\mu=0) \sim V \mu^2 T^2 \sim 1$ as was discussed 
in detail in \cite{1loop}.

For $\mu > m_\pi/2$ a Bose Einstein condensate of pions forms in the phase
quenched theory and the mean field terms in the chiral Lagrangian contribute 
to $\langle \exp(2i\theta')\rangle$. These terms are of order $\mu^2 F^2 V \sim
V/L^2 \sim L^2$. Hence, for $\mu > m_\pi/2$, the strength of the sign problem
depends on $L$ even if we scale $m_\pi$ and $\mu$ with $L$ according to 
$p$-counting.

Since the difference of the finite parts of the one-loop free energy
appears repeatedly below, it will be convenient to introduce the notation  
\be
\Delta G_0 \equiv \Delta G_0(\mu,-\mu,m,m) \equiv
G_0(\mu,-\mu,m,m)-G_0(\mu,\mu,m,m) = g_0(\mu,-\mu,m,m)-g_0(\mu,\mu,m,m).    
\label{DG}
\ee

Below we will also meet free energies where the chemical potentials are
not of opposite sign and where the quark masses are different. To be
precise we reserve the 
notation $\Delta G_0$ as defined in (\ref{DG}), and explicitly write the
dependence on the chemical potentials and quark masses when necessary.   

\section{The distribution of the phase \hspace{5mm} ($\mu<m_\pi/2$)}
\label{sec:thdistmu<0p5mpi}

The distribution of the phase angle can be 
obtained from the moments of the phase factor \cite{1loop}
\be\label{thdist-moments-sum}
\langle\delta(\theta-\theta')\rangle_{N_f}
=\frac{1}{2\pi} \sum_{p=-\infty}^\infty e^{-ip\theta}\langle
e^{ip\theta'}\rangle_{N_f}. 
\ee
The even moments are ratios of  a partition function with $p$ additional determinants and
inverse conjugate determinants and the usual $N_f$ flavor partition function
\be
\langle e^{2ip\theta'}\rangle_{N_f} =\frac{1}{Z_{N_f}}\left \langle 
 \frac{{\det}^p(D +\mu\gamma_0+m)}
{{\det}^p(D-\mu\gamma_0+m)} {\det}^{N_f}(D+\mu \gamma_0 +m)\right \rangle.
\label{mom-basic}
\ee
Since the number of charged Goldstone modes of the partition function
in the numerator is
$p(p+N_f)$ whereas the contributions of the neutral Goldstone bosons from
the numerator and the denominator cancel, we obtain  
\be
\langle e^{2ip\theta'}\rangle_{N_f} &=& e^{-p(N_f+p)\Delta G_0}.
\label{exp2nith}
\ee
When the quark mass is outside the support of the Dirac spectrum, 
the contribution to the phase angle of individual eigenvalues is in the
range $ [-\pi/2, \pi/2] $, and we expect half-integer powers of
the determinants in (\ref{mom-basic}) are smoothly connected to results
obtained for integer powers. In other words, we expect that
the replica trick \cite{Edwards,replica} can be used 
to analytically continue the moments to half integer values of
$p$. We then find
\be
\langle\delta(\theta-\theta')\rangle_{N_f}
&=& \frac 1{2\pi} \sum_{p=-\infty}^\infty e^{-ip\theta-(p/2)((p/2)+N_f) 
\Delta G_0} \nn \\
&=& \frac 1{2\pi}e^{iN_f\theta +\frac 14 N_f^2 \Delta G_0} 
\sum_{u=-\infty}^\infty e^{-iu\theta-u^2 \Delta G_0/4}\nn \\
&=& \frac 1{2\pi}e^{iN_f\theta +\frac 14 N_f^2 \Delta G_0}
\vartheta_3(\theta/(2\pi), e^{-\Delta G_0/4}) .
\label{deltanf}
\ee
After a Poisson resummation this can be rewritten  as \cite{1loop}
\be
\langle\delta(\theta-\theta')\rangle_{N_f} 
&=& \frac 1{\sqrt{\pi \Delta G_0}} e^{iN_f\theta +\frac 14 N_f^2 \Delta G_0}
\sum_{n=-\infty}^\infty e^{-(\theta+2n\pi )^2/\Delta G_0},
\qquad \theta\in[-\pi,\pi]
\label{rhoth-CPT}
\ee
valid for a compact phase angle $\theta\in[-\pi,\pi]$.

\begin{figure}[t!]
  \unitlength1.0cm
  \epsfig{file=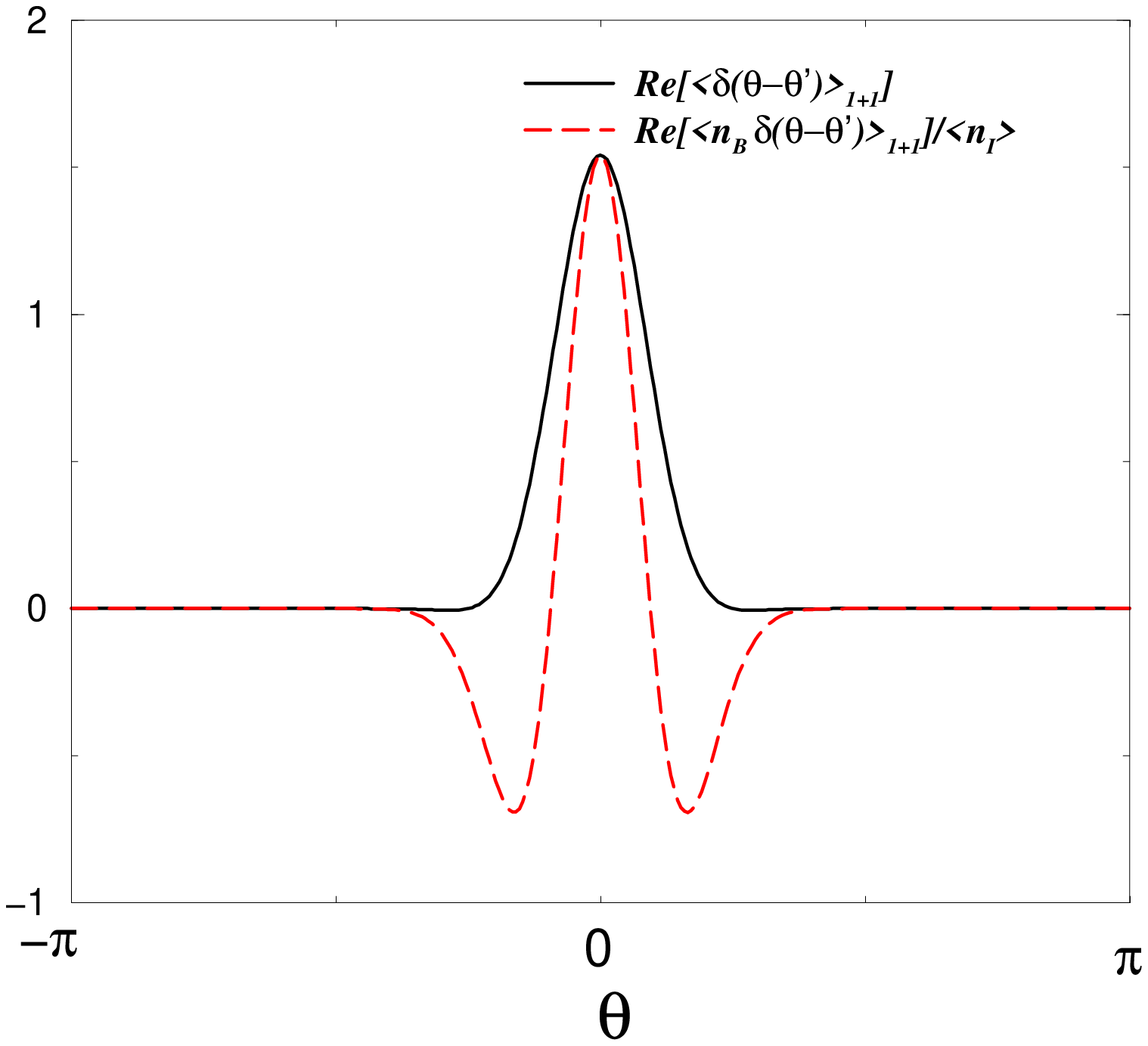,clip=,width=8cm} \hspace{3mm}
 \epsfig{file=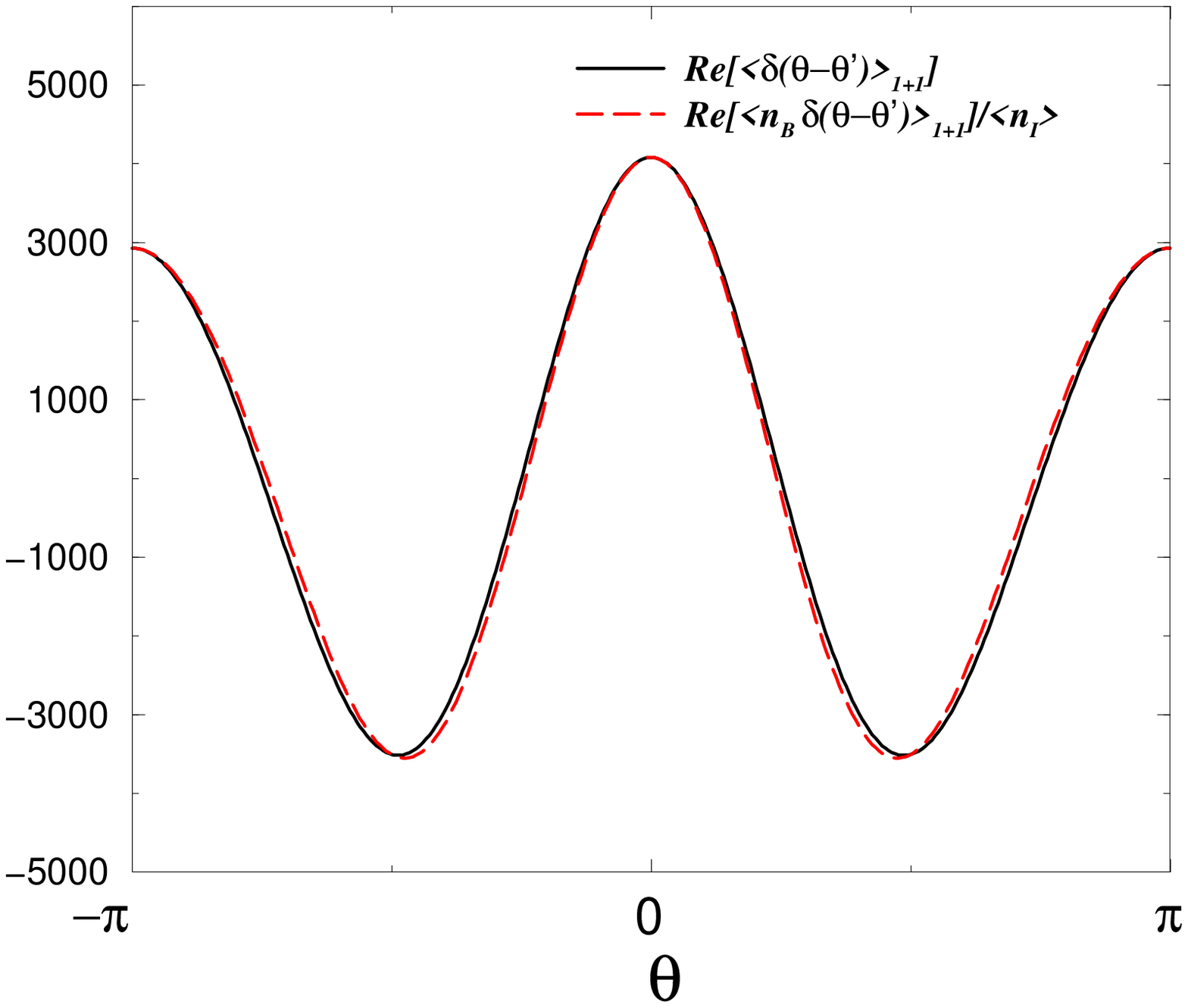,clip=,width=8.4cm}
  \caption{\label{fig:nBth-dist} The real part of the distribution of the
    phase $\langle\delta(\theta-\theta')\rangle_{1+1}$ (solid curve) 
    for $\Delta G_0=0.2$ left and $\Delta G_0=10$ right. Also shown is the real
    part of the distribution of the 
    baryon number over $\theta$ (dashed curve). For better comparison the
    latter has been rescaled by $(\lim_{\tilde{\mu}\to\mu}
    \frac{d}{d\tilde{\mu}} \Delta G_0(-\mu,\tilde{\mu}))$. The fact that the
    $\theta$-distribution is 
    normalized to unity while the distribution of the baryon number over
    $\theta$ integrates to zero is not easy to see when $\Delta G_0 =10$. 
    This directly illustrates the severity of the sign problem. Note
    that the phase is constrained to $\theta\in[-\pi,\pi]$.} 
\end{figure}

Notice that
\be
\frac{Z_{N_f}} {Z_{|N_f|}} = e^{-\frac 14 N_f^2 \Delta G_0}
\ee
so that to be consistent with the general form given in 
(\ref{th-distBasic}),
the result (\ref{rhoth-CPT}) shows that the quenched and the phase-quenched
$\theta$-distributions are identical. Also note that the
$\theta$-distribution depends only on $\Delta G_0$. 
Plots for $\Delta G_0=0.2$ and $\Delta G_0=10$ are shown in figure
\ref{fig:nBth-dist}. Notice the different scales in the two plots. For
$\Delta G_0 =10$, when the sign problem is severe, the normalization to
one requires a delicate cancellation.  

\vspace{2mm}

As long as the contribution to the phase of the fermion determinant  from 
individual eigenvalue
pairs does not exceed $\pi/2$ one can unambiguously define the phase of the
determinant on $[-\infty,\infty]$ as was done by Ejiri \cite{Ejiri}. To obtain
this distribution simply interpret  the angle in (\ref{rhoth-CPT}) as
ranging from $-\infty$ to $\infty$. This leads to the 
Gaussian distribution (here for $N_f = 2$)
\be
\hspace{1cm}
\langle\delta(\theta-\theta')\rangle_{1+1} = \frac{e^{2i\theta}}{\sqrt{\pi
    \Delta G_0}}e^{-\theta^2/\Delta G_0+\Delta G_0}, \qquad 
\theta\in[-\infty,\infty].
\label{th-dist}
\ee
However, when the quark mass is inside the support of the
spectrum of the Dirac operator 
only the phase restricted $[-\pi,\pi]$ can be defined
uniquely. We return to this point in section \ref{sec:Lorentzian} where we
derive the $\theta$-distribution for $\mu>m_\pi/2$.

When the angles are noncompact and
replica trick can be used, it is useful to represent 
the $\delta$-function in Eq. (\ref{deltanf})
by an integral over $p$ instead of a sum over $p$. Below 
this will be exploited on several occasions to simplify our expressions.

\section{The baryon number operator ($\mu<m_\pi/2$)} 
\label{sec:baryon}

Since the pions have zero baryon charge the baryon number in chiral
perturbation theory is
automatically zero. We will see below that
the baryon number at fixed $\theta$ is a total derivative.

\vspace{3mm}

To derive $\langle n_B\delta(\theta-\theta')\rangle$ we first compute the
correlation between the baryon number and all moments of the phase factor
\be
\langle n_B \ e^{2ip\theta'}\rangle_{1+1} 
= \frac{1}{2Z_{1+1}}\lim_{\tilde{\mu}\to\mu} \frac{d}{d\tilde{\mu}} 
\langle \frac{{\det}^p(D +\mu\gamma_0+m)}
{{\det}^p(D-\mu\gamma_0+m)} {\det}^2(D+\tilde{\mu} \gamma_0 +m) \rangle.
\ee
To one-loop order in chiral perturbation theory we obtain 
\be
\frac{\langle \frac{{\det}^p(D +\mu\gamma_0+m)}
{{\det}^p(D-\mu\gamma_0+m)} {\det}^2(D+\tilde{\mu} \gamma_0 +m) \rangle}
{\langle {\det}^2(D+\mu \gamma_0 +m)  \rangle} = 
e^{-2p(\Delta G_0(-\mu,\tilde{\mu})-\Delta G_0(\mu,\tilde{\mu}))-p^2 \Delta G_0(-\mu,\mu)}.
\ee
To keep track of the combinatorics it is essential to recall that the 
one-loop free energy does not depend on the baryon chemical potential, that is
$G_0(\mu,\mu)=G_0(\mu=0)$. We conclude that 
\be
\langle n_B \ e^{2ip\theta'}\rangle_{1+1} = -\left (
\lim_{\tilde{\mu}\to\mu} \frac{d}{d\tilde{\mu}} \  p \ \Delta G_0(-\mu,\tilde{\mu})\right )
e^{-p(2+p)\Delta  G_0(-\mu,\mu)}.
\label{nBexp2ipth} 
\ee
The delta function $\delta(\theta - \theta') $ is obtained after summing over
$p$. Interpreting the phase angle on $\langle -\infty, \infty \rangle$ and 
proceeding in the same way as for the distribution of $\theta $ we obtain
\be
\langle n_B \ \delta(\theta-\theta')\rangle_{1+1} 
 =  \left(\lim_{\tilde{\mu}\to\mu} \frac{d}{d\tilde{\mu}} \Delta G_0(-\mu,\tilde{\mu})\right)(1+i\frac{\theta}{\Delta G_0})\frac{e^{2i\theta}}{\sqrt{\pi \Delta
    G_0}}e^{-\theta^2/\Delta G_0+\Delta G_0}.
\ee
The total baryon number density should vanish because chiral perturbation
theory does not include baryonic degrees of freedom. This can be seen simply by
writing the above expression as a total derivative
\be
\langle n_B \rangle_{1+1} =\left(\lim_{\tilde{\mu}\to\mu} \frac{d}{d\tilde{\mu}} \Delta
  G_0(-\mu,\tilde{\mu})\right) \frac{1}{\sqrt{\pi \Delta G_0}} 
\int_{-\infty}^\infty d\theta  \frac 1{2i}\frac d{d\theta}
e^{2i\theta}e^{-\theta^2/\Delta G_0+\Delta G_0} = 0.  
\label{nB0}
\ee
The total derivawtive appears because all one-loop contributions to
the $2p$'th moment of the phase factor are proportional to $p$ or $p^2$ so
that the differentiation to obtain the baryon density leads to an 
overall factor $p$. This factor can be expressed as a total derivative with
respect to $\theta$. 
Notice that when $\Delta G_0 \gg 1$ the extreme tail of the distribution over
$\theta$ can contribute significantly to the cancellation of the total baryon
number.

If we, as is usually the case,  
consider the phase on $[-\pi,\pi]$ we get instead 
\be
\langle n_B \ \delta(\theta-\theta')\rangle_{1+1} 
 =  \left(\lim_{\tilde{\mu}\to\mu} \frac{d}{d\tilde{\mu}} \Delta
   G_0(-\mu,\tilde{\mu})\right)\sum_{n=-\infty}^{\infty}(1+i\frac{\theta+2\pi
   n}{\Delta G_0})\frac{e^{2i\theta}}{\sqrt{\pi \Delta
    G_0}}e^{-(\theta+2\pi n)^2/\Delta G_0+\Delta G_0}.
\ee 
 An illustration of $\langle n_B \
\delta(\theta-\theta')\rangle_{1+1}$ is given in figure
\ref{fig:nBth-dist}. For small $\Delta G_0$, a small phase angle gives an
excess of baryons over anti-baryons, which is cancelled by the opposite
effect at larger phase angle, resulting in $n_B =0$.
For large $\Delta G_0$ the plot is quite similar to the
$\theta$-distribution which is 
also shown in this figure. There is however an important 
difference: The integral over $\theta$ of the $\theta$-distribution
is unity while the total baryon number is zero.

The importance of the tail for the cancellation of the total baryon number
translates into the importance of the terms with large values of $|n|$.

\section{The off-diagonal susceptibility ($\mu<m_\pi/2$)} 
\label{sec:baryonSQ}

Even though pions have zero baryon charge chiral perturbation theory gives a
nontrivial prediction for the off-diagonal quark number susceptibility. To
compute this expectation value we start from
\be
Z_{1+1}(\mu,\mu_a) = \langle {\det}(D+\mu \gamma_0 +m){\det}(D+\mu_a \gamma_0
+m) \rangle.  
\ee
where $\langle \ldots \rangle$ is the quenched average. The average of the
off-diagonal susceptibility is then given by
\be
\langle \chi \rangle_{1+1} 
= \frac{1}{Z_{1+1}(\mu,\mu)} \lim_{\mu_a\to\mu} 
\frac{d}{d\mu}\frac{d}{d\mu_a} Z(\mu,\mu_a).
\ee
To one-loop order in chiral perturbation theory we find
\be
\frac{Z_{1+1}(\mu_a,\mu_b)}{Z_{1+1}(\mu,\mu)} 
= e^{G_0(\mu_a,\mu_b)-G_0(\mu=0)}. 
\ee
The one loop contribution $G_0(\mu_a,\mu_b)$ to the free energy
from a charged pion pair made out of quarks with chemical potentials $\mu_a$
and $\mu_b$ only depends on the absolute value of the difference
$\mu_a-\mu_b$. Moreover, since $\lim_{\mu_a\to\mu}d/d\mu G_0(\mu,\mu_a)=0$ we
immediately get 
\be
\langle \chi \rangle_{1+1} 
= \lim_{\mu_a\to\mu} \frac{d}{d\mu}\frac{d}{d\mu_a} \Delta G_0(\mu,\mu_a).
\label{nbSQ}
\ee

\subsection{The distribution}

To compute the contribution of configurations
with a specific phase to the off-diagonal susceptibility  
we first compute  the moments $\langle  \chi \
e^{2ip\theta'}\rangle_{1+1}$.  We start from 
\be
Z_{1+1+p|p^*}(\mu_a,\mu_b,\mu|\mu)
= \langle\frac{{\det}^p(D +\mu\gamma_0+m)}
{{\det}^p(D-\mu\gamma_0+m)} 
{\det}(D+\mu_a \gamma_0 +m){\det}(D+\mu_b \gamma_0
+m) \rangle,  
\label{Zsusy_nSQ}
\ee 
and evaluate the limit
\be
\langle \chi \ e^{2ip\theta'}  \rangle_{1+1} 
= \frac{1}{Z_{1+1}(\mu,\mu)}\lim_{\mu_a,\mu_b\to\mu}
\frac{d}{d\mu_a}\frac{d}{d\mu_b}
Z_{1+1+p|p^*}(\mu_a,\mu_b,\mu|\mu).  
\ee
For the fermionic Goldstone modes we have an additional minus sign leading to
\be
\frac{Z_{1+1+p|p^*}(\mu_a,\mu_b,\mu|\mu)}{Z_{1+1}(\mu,\mu)}
= e^{-p\Delta G_0(\mu_a,-\mu)-p\Delta G_0(\mu_b,-\mu)-p^2\Delta
  G_0(\mu,-\mu)+p\Delta 
  G_0(\mu_a,\mu)+p\Delta G_0(\mu_b,\mu)+\Delta
  G_0(\mu_a,\mu_b)}\, . 
\ee
Keeping track of $p$ we find
\be
\langle  \chi \ e^{2ip\theta'} \rangle_{1+1} 
= \lim_{\mu_a\to\mu} \left[p^2 (\frac{d}{d\mu} \Delta
  G_0(\mu_a,-\mu))^2+\frac{d}{d\mu} \frac{d}{d\mu_a} \Delta G_0(\mu_a,\mu) \right]
e^{-p(2+p)\Delta G_0(\mu,-\mu)}.
\label{susp2}
\ee
For a non-compact phase angle $\theta\in[-\infty,\infty]$ we obtain a 
$\delta$-function in the l.h.s. after integrating over $p$. Proceeding in the
same way as for the distribution function of the phase we find 
\be
\langle \chi \ \delta(\theta-\theta')  \rangle_{1+1} = \left([(1+i \frac{\theta}{\Delta G_0})^2+\frac{1}{2\Delta G_0}][\frac{d}{d\mu}
  \Delta  G_0(\mu_a,-\mu)]^2_{\mu_a=\mu}+\frac{d}{d\mu} \frac{d}{d\mu_a} \Delta
  G_0(\mu_a,\mu)_{\mu_a=\mu} \right)\frac{e^{2i\theta}}{\sqrt{\pi \Delta
    G_0}}e^{-\theta^2/\Delta G_0+\Delta G_0} .\nn \\
\ee  
The first term between round brackets results from the term $\sim p^2$
in Eq. (\ref{susp2}) which, before summing over $p$ can be simply rewritten
as second derivative with respect to $\theta$,
\be
 [\frac{d}{d\mu}
  \Delta  G_0(\mu_a,-\mu)]^2_{\mu_a=\mu}
 \frac 1{(2i)^2} \frac {d^2}{d\theta^2}
\frac{e^{2i\theta}}{\sqrt{\pi \Delta G_0}}e^{-\theta^2/\Delta G_0+\Delta G_0} .
\ee   
and vanishes upon integration over $\theta$. The $\theta$ dependence 
of the second term is the same as for
the $\theta$-distribution which is normalized to 1.
Upon integration over
the angle $\theta$ we  
thus  recover the expectation value of the susceptibility (\ref{nbSQ}). 
Again we emphasize that for $\Delta G_0 \gg 1$ contributions from 
the extreme tail may give important contributions to the off-diagonal quark
number susceptibility.

\section{The chiral condensate ($\mu<m_\pi/2$)} 
\label{sec:Sigma}

In this section we compute the chiral condensate when the phase angle 
of the fermion determinant is constrained to $\theta$. This quantity is defined
as
\be
\langle
\bar \psi \psi \ \delta(\theta -\theta') \rangle.
\ee
Since the chiral condensate is nonzero for $T=0$ and $\mu=0$
this derivation requires also the divergent part of the free energy. 
The required generating functional has different masses    
\be
\frac{\langle\frac{{\det}^p(D +\mu\gamma_0+m)}
{\det^p(D-\mu\gamma_0+m)} {\det}(D+\mu\gamma_0 +\tilde{m})^2 \rangle}{\langle
{\det}(D+\mu\gamma_0 +m)^2\rangle}.   
\ee 
The desired expectation value $\langle e^{2 i p\theta'} \bar\psi\psi
\rangle_{1+1}$ is obtained by taking the derivative with respect to
 $\tilde{m}$ and
subsequently the limit $\tilde{m}\to m$. The combinatorics is much like for the
baryon number, but  here we have to keep track of both mass derivatives
and the chemical potentials. This leads to  
\be
\label{exp2ipthSigma}
\langle \bar\psi\psi \  e^{2 i p\theta'}\rangle_{1+1} 
& = & \frac{1}{2}\lim_{\tilde{m}\to m} \frac{d}{d\tilde{m}} \frac{\langle\frac{{\det}^p(D +\mu\gamma_0+m)}
{\det^p(D-\mu\gamma_0+m)} {\det}(D+\mu\gamma_0 +\tilde{m})^2 \rangle}{\langle
{\det}(D+\mu\gamma_0 +m)^2\rangle} \\
& = & \big(\langle \bar\psi\psi\rangle_{1+1}^{0} +\frac{d}{d\tilde{m}}
\Big[-p(G_0(\mu,-\mu,\tilde{m},m)-G_0(\mu,\mu,\tilde{m},m))\nn\\
&& \hspace{3.5cm} +(G_0(\mu,\mu,\tilde{m},\tilde{m})-G_0(\mu,\mu,m,m)) \Big]_{\tilde{m}=m}
\big)e^{-p(2+p)\Delta G_0}.\nn
\ee
Where $\langle \bar\psi\psi\rangle_{1+1}^{0}$ is the one-loop
renormalized chiral condensate at zero temperature and zero chemical potential.  
For $p=0$ we obtain the one-loop renormalized chiral condensate at nonzero 
temperature and nonzero chemical potential
\be
\label{SigmaFull}
\langle \bar\psi\psi \rangle_{1+1} 
& = & \langle \bar\psi\psi\rangle_{1+1}^{0} +\frac{d}{d\tilde{m}}
\Big[(G_0(\mu,\mu,\tilde{m},\tilde{m})-G_0(\mu,\mu,m,m)) \Big]_{\tilde{m}=m},
\ee
which is independent of the chemical potential.

\vspace{2mm}

The distribution of the chiral condensate over the phase $\theta$ is obtained
after multiplication by $\exp(-ip\theta)$ and integrating over $p$
\be 
\langle  \bar\psi\psi \ \delta(\theta-\theta') \rangle_{1+1} & = &
\Big(\langle \bar\psi\psi\rangle_{1+1}^{0}
+\lim_{\tilde{m}\to m} \frac{d}{d\tilde{m}}
\Big[(1+i\frac{\theta}{\Delta
    G_0})(G_0(\mu,-\mu,\tilde{m},m)-G_0(\mu,\mu,\tilde{m},m)) \nn \\
&& \hspace{2cm} +(G_0(\mu,\mu,\tilde{m},\tilde{m})-G_0(\mu,\mu,m,m))\Big] \Big) \frac{e^{2i\theta}}{\sqrt{\pi \Delta
    G_0}}e^{-\theta^2/\Delta G_0+\Delta G_0}.
\label{Sigma-dist}
\ee
The factor $1+i\theta/\Delta G_0$ can again be written as a total derivative
of the exponential factors. Upon integration the contribution from this term
vanishes, and
we recover the full
condensate (\ref{SigmaFull}) 
\be
\langle  \bar\psi\psi \rangle_{1+1} & = & \int_{-\infty}^\infty d\theta \
\langle \bar\psi\psi \ \delta(\theta-\theta') \rangle_{1+1}. 
\ee
Again important tail contributions arise for $\Delta G_0 \gg 1$.

\section{The $\theta$-distribution for an ensemble generated at $\mu=0$}
\label{sec:gen@mu0}

In the method of Ejiri \cite{Ejiri} one evaluates the $\theta$-distribution
as a function of the chemical potential for an ensemble generated at zero
chemical potential. Here we compute this partially quenched
$\theta$-distribution within one-loop chiral perturbation theory.

\vspace{2mm}

We start out evaluating the moments of the phase factor for an ensemble
generated at zero chemical potential
\be
&&\frac{1}{Z_{1+1}(\mu=0)}\left\langle\frac{{\det}^p(D+\mu\gamma_0+m)}{{\det}^p(D-\mu\gamma_0+m)}
{\det}^2(D+m) \right  \rangle = e^{-p^2 \Delta G_0(\mu,-\mu)}.
\ee
We then obtain the distribution (for $\mu<m_\pi/2$)
\be
\langle\delta(\theta-\theta')\rangle_{\mu=0} 
= \frac{1}{\sqrt{\pi\Delta G_0}}e^{-\frac{\theta^2}{\Delta G_0}}.
\ee
This one-loop prediction is identical to that for the quenched and phase
quenched ensemble: Whether we compute the width of the Gaussian
for the $\theta$-distribution in the full ensemble generated at $\mu$, or the
full ensemble generated at $\mu=0$, or in the quenched ensemble, or the 
phase quenched ensemble, we find exactly the same result. 

\vspace{3mm}

Ejiri also has studied distributions of  
$F=|\det(D+\mu\gamma_0+m)|/\det(D+m)$. His assumption is
that the $\theta$ distribution remains Gaussian even for a fixed value of
$F$. As we shall see below, this assumption is justified for
$\mu<m_\pi/2$ to one-loop order in chiral
perturbation theory.

\section{The $\epsilon$-regime}
\label{sec:epsilon}

The above analysis  
suggests that the chemical potential has to be of the
order of $1/\sqrt{V}$  to suppress the correlation between the phase
and the chiral condensate or baryon density. Such a scaling corresponds
to the $\epsilon$-regime \cite{VS,V} where the dimensionless quantities 
\be
\hat{m}\equiv m \Sigma V\ \ \ \ {\rm and}  \ \ \ \ \hat\mu^2\equiv\mu^2 F^2 V, 
\ee
are kept fixed for $V\to\infty$. Here and below $\Sigma$ and $F$ are the
chiral condensate and the pion decay constant as they appear in the chiral
Lagrangian.   
Note that it is possible to go smoothly between the $\epsilon$- and
$p$-regime see \cite{DF}.
 
In the $\epsilon$-regime the moments of the phase factor remain finite for
$V\to\infty$ \cite{exp2ith-letter,phase-long} 
\be 
\langle e^{2i p\theta'}\rangle_{N_f} = (1-2\hat\mu^2/\hat{m})^{p(p+N_f)},
\ee
where we quote the result valid for $\hat{m},\hat\mu\gg1$ and
$2\hat\mu^2<\hat{m}$. 
To obtain the distribution of the phase in the $\epsilon$-regime let us 
rewrite this as
\be 
\langle e^{2i p\theta'}\rangle_{N_f} = e^{-p(p+N_f) \Delta \hat{G}_0},
\ee
with
\be
\Delta \hat{G}_0 = - \log(1-2\hat\mu^2/\hat{m}).
\ee
It is of exactly the same form as Eq.~(\ref{exp2nith}).
So we again find the distribution (\ref{th-dist}) but now with $\Delta
\hat{G}_0$ instead of $\Delta G_0$
\be
\langle\delta(\theta-\theta')\rangle_{1+1} = \frac{e^{2i\theta}}{\sqrt{\pi
    \Delta \hat{G}_0}}e^{-\theta^2/\Delta \hat{G}_0+\Delta \hat{G}_0}.
\label{th-dist-eps}
\ee
The variance of the Gaussian envelope starts out
at zero for small $\mu$ (i.e. for $2\hat\mu^2 \ll \hat{m}$) and approaches
infinity as $\log(1-x^2)$ for $x\to1$.

\newpage

\section{The $\theta$-distribution for $\mu>m_\pi/2$}
\label{sec:Lorentzian}

We now turn to the distribution of the phase of the fermion determinant
when the quark mass is inside the support of the
Dirac operator. For low $T$ this means that $\mu > m_\pi/2$.  

Because the Dirac operator is only determined up to an operator with
determinant equal to unity,
\be
\det[D +\mu\gamma_0+m] = \det[A(D+\mu\gamma_0 +m)] \quad {\rm with}
\quad \det A =1,
\ee
the sum of the phases of individual eigenvalues of the 
Dirac operator may differ by multiples or $2\pi$
depending on the choice of $A$.
When the quark mass is outside the eigenvalue distribution the contribution
to the phase of
a pair $\lambda+m,-\lambda+m$ of eigenvalues is less than $\pi/2$, 
and in this case the sum of the phases of all the eigenvalues does not depend
on $A$. Therefore it makes sense 
to extend the total phase to $\langle -\infty, \infty \rangle$. 
For $\mu > m_\pi/2$ 
the phase of the determinant, however, differs by multiples of $2\pi$ depending
on the choice of $A$. Therefore, when $\mu > m_\pi/2$, 
it only makes sense to define the phase modulo $2\pi$.

\vspace{2mm}

As before, the $\delta$-function, $\delta(\theta-\theta')$, will be obtained
from the moments of the 
phase factor which now are dominated by the leading order term in the chiral
expansion.
Not surprisingly, this leads to a much wider $\theta$-distribution. What is
perhaps somewhat surprising is that, as will be shown below, the distribution 
now takes a Lorentzian shape.  Because of ambiguities in the phase angle
we do not expect that we can use the replica trick to calculate half-integer
moments of the phase factor. Therefore we will only evaluate the even
moments, $\langle \exp(2ip\theta') \rangle$, with integer values of $p$. This
is sufficient to obtain the full distribution of the total phase angle,
$2\theta \in [-\pi,\pi]$, of $\det(D+\mu\gamma_0+m)^2$ relevant for the two
flavor theory 
\be
\langle \delta(2\theta -2\theta')\rangle =
\frac 1{\pi} \sum_{p=-\infty}^\infty e^{-2ip\theta} 
\langle e^{2ip\theta'} \rangle.
\label{evendist}
\ee
Alternatively this can be seen as the combination $\frac 12[ \langle
\delta(\theta -\theta')\rangle + \langle \delta(\theta -\theta'+\pi)\rangle]$
of the distribution of the phase angle, $\theta$, of $\det(D+\mu\gamma_0+m)$.

\subsection{Bosonic partition function}
\label{subsec:bosMF}

The moments of the phase factor involve inverse powers of determinants,
c.f. Eq.~(\ref{mom-basic}). As was realized when investigating 
the partition function
with one bosonic flavor such inverse determinants lead to a phase transition
at $\mu=m_\pi/2$.   
In order to compute the moments of the phase factor for $\mu>m_\pi/2$ to
leading order in chiral perturbation theory we therefore first recall the
explanation of  the exact results for the bosonic partition
function (obtained by  integration over the Goldstone manifold
\cite{SVZ} or from the Cauchy transform of the fermionic partition function
\cite{AP,Bergere,SV-bosonic}) in terms of a mean field argument.

The observation of \cite{SV-bosonic} is that the bosonic partition function 
\be
\left \langle \frac{1}{\det(D+\mu\gamma_0+m)}\right \rangle = 
\left \langle \frac{\det(D-\mu\gamma_0+m)}{\det(D+\mu\gamma_0+m)(D-\mu\gamma_0+m)}\right\rangle 
\ee
at a mean field level behaves like 
\be
\frac{\langle\det(D-\mu\gamma_0+m)\rangle}{\langle\det(D+\mu\gamma_0+m)(D-\mu\gamma_0+m)\rangle
} .
\label{MF-Zm1}
\ee
The reason is loosely speaking that the inverse determinant must be
regularized in order to be convergent and that Grassmannian mean
field terms are absent.   

The denominator of Eq.~(\ref{MF-Zm1}) is the phase quenched theory which has a
phase transition at $\mu=m_\pi/2$. The mean
field result for the phase quenched theory is given by
\be
\langle\det(D+\mu\gamma_0+m)(D-\mu\gamma_0+m)\rangle = e^{-V L_I},
\ee
where \cite{KSTVZ,SS}
\be
L_I= -2\mu^2F^2 -\frac{\Sigma^2 m^2}{2\mu^2F^2}
\label{LI}
\ee
is the static Lagrangian for $\mu > m_\pi/2$. The average of the determinant
in the numerator of Eq.~(\ref{MF-Zm1}) is the familiar one flavor partition
function (which is independent of $\mu$ in chiral perturbation theory)  
\be
\langle\det(D-\mu\gamma_0+m)\rangle =e^{-V L_0/2},
\label{Zfermionic}
\ee
where 
\be
L_0 = - 2 m \Sigma
\ee 
is the mean field Lagrangian at $\mu =0$. 
In conclusion, the mean field result for the bosonic partition function 
is given by
\be
\langle \frac{1}{\det(D+\mu\gamma_0+m)}\rangle = e^{-V L_0/2 + V L_I}.
\ee
As shown in detail in \cite{SV-bosonic,SVZ} this gives the correct mean
field physics. Note the striking difference with the fermionic partition
function (\ref{Zfermionic}) which is independent of the chemical potential.

\subsection{The quenched $\theta$-distribution}

Let us now use what we learned from the bosonic case to compute 
the quenched distribution of the phase of the fermion determinant
for $\mu>m_\pi/2$: We will first show
that 
\be
\langle e^{2ip\theta'}\rangle = e^{- V {L_B} |p|},
\label{momQ_2mu>mpi}
\ee
where $L_B = L_0 - L_I$ with $L_0$ and $L_I$ given above (note that
$L_B\geq0$).  

Since by charge conjugation symmetry $\langle \exp(2ip\theta')\rangle=\langle
\exp(-2ip\theta')\rangle$ this expectation value only depends on the
absolute value of $p$, and we only need to
consider $p>0$. 
First, we rewrite the moments as
\be
 \left  \langle e^{2ip\theta'}\right \rangle = \left
\langle\frac{(\det(D+\mu\gamma_0+m) \det(D+\mu\gamma_0+m))^{p}}{(\det(D+\mu\gamma_0+m) \ {\det}(D-\mu\gamma_0+m))^{p}}
\right\rangle.
\ee
Now the contribution from the denominator is the inverse of 
the replicated phase quenched theory. This was worked out in \cite{SpecPhase}
\be
\frac{1}{\langle (\det(D+\mu\gamma_0+m) {\det}(D-\mu\gamma_0+m))^{p}\rangle} = e^{{p} V L_I}.
\ee
The contribution from the numerator is just 
\be
\langle (\det(D+\mu\gamma_0+m) \ \det(D+\mu\gamma_0+m))^{p} \rangle 
= e^{-{p} V L_0},
\label{MFreplicatedZpq}
\ee
which together with the previous result reproduces Eq.~(\ref{momQ_2mu>mpi}). The
sum over $p$ results in 
\be\label{thdist-Qmu>mpiO2}
\langle \delta(2\theta-2\theta')\rangle 
&=& \frac{1}{\pi} \sum_{p=-\infty}^\infty \ e^{-2i \theta p}e^{- V {L_B}
  |p|}\nn\\ 
&=& \frac{1}{\pi}\sum_{n=-\infty}^\infty\frac{2V L_B}{(V L_B)^2+(2\theta+2\pi n)^2}. 
\ee
The sum  over $n$ can be evaluated as
\be\label{thdist-Qmu>mpiO2v2}
\langle \delta(2\theta-2\theta')\rangle
= \frac{1}{\pi}\frac{\sinh(V L_B)}{\cosh(V L_B)-\cos(2\theta)} \ .
\ee
This is a compactified Lorentzian, centered at zero. We recall that
$2\theta\in[-\pi,\pi]$ is the phase of $\det(D+\mu\gamma_0+m)^2$.

\subsection{The unquenched $\theta$-distribution}

To calculate the unquenched $\theta$-distribution function we again consider
the moments $\langle \exp(2ip\theta') \rangle_{N_f}$. They can be rewritten as
\be
\langle e^{2ip\theta'}\rangle_{N_f} = \frac{1}{Z_{N_f}} \langle
\frac{\det^{u}(D+\mu\gamma_0+m)}{\det^{* \ u}(D+\mu\gamma_0+m)} ({\det}^* (D+\mu\gamma_0+m)\det(D+\mu\gamma_0+m))^{N_f/2} \rangle, 
\ee
where we have introduced $u=p+N_f/2$.  
By charge conjugation, this expectation values does not
 depend on the sign of $u=p+N_f/2$, i.e. it only depends on $|u|$,
and it only has to be calculated  for $p\ge -N_f/2$. We separately consider
the cases $p >0$ and $0\ge  p \ge -N_f/2$.

For $p > 0$ there are inverse powers of ${\det}^*$, and we apply the rules of
section \ref{subsec:bosMF} 
\be
\frac{1}{Z_{N_f}} \langle e^{2ip\theta'}{\det}^{N_f}(D+\mu\gamma_0+m)\rangle
&=&  \frac{1}{Z_{N_f}}\left \langle \frac{\det^{2p+N_f}(D+\mu\gamma_0+m)}{(\det(D+\mu\gamma_0+m) \ {\det}(D-\mu\gamma_0+m))^{p}}
\right \rangle \\ 
&\simeq&  \frac{1}{Z_{N_f}}
\frac{\langle\det^{2p+N_f}(D+\mu\gamma_0+m)\rangle}{\langle(\det(D+\mu\gamma_0+m)
  \ {\det}(D-\mu\gamma_0+m))^{p}\rangle}, \nn 
\ee
where the final equality holds at the mean field level.
The contribution from the denominator follows again from the result
of the replicated fermionic theory $\langle(\det \
{\det}^*)^{p}\rangle$, see Eq.~(\ref{MFreplicatedZpq}). The 
numerator is equal to  $\exp(-(p+N_f/2) \ L_0)$ and the normalization, 
$1/Z_{N_f}$, gives $\exp(N_f/2 \ L_0)$.  
Therefore the $N_f$ dependence cancels, 
and we find the quenched result for $p > 0$
\be
\langle e^{2ip\theta'}\rangle_{N_f} = e^{-p V L_B},
\hspace{2cm} p\geq0.
\label{pgeq0}
\ee
Here, we extended the equality to $p = 0$ which is satisfied trivially.

\vspace{3mm}

Now, let us look at negative values of $p$. This means that $\det^*$ is in the
numerator and $\det$ is in the denominator. 
For  $-N_f/2 \leq p \leq 0$  the moments can be rewritten as
\be
 & & \frac{1}{Z_{N_f}} \langle \frac{\det^{p}(D+\mu\gamma_0+m)}{\det^{* \ p}(D+\mu\gamma_0+m)}
 {\det}^{N_f}(D+\mu\gamma_0+m)\rangle                   
\nn\\
 & = &  \frac{1}{Z_{N_f}} \langle \frac{\det^{* \ |p|}(D+\mu\gamma_0+m)}
{\det^{|p|}(D+\mu\gamma_0+m)}
 {\det}^{N_f}(D+\mu\gamma_0+m)\rangle ,
\nn\\
 & = &
    \frac{1}{Z_{N_f}} \langle
    (\det(D+\mu\gamma_0+m){\det}^{*}(D+\mu\gamma_0+m))^{|p|}
    {\det}^{N_f-2|p|}(D+\mu\gamma_0+m) \rangle .
\ee
Note that both exponents are positive. The $|p|$ pairs of conjugate quarks
form a pion condensate while, at the mean field level, 
the $N_f-|p|$ quarks are passive spectators resulting 
in the average phase factor
\be
\langle e^{2ip\theta'}\rangle _{N_f}
& = & e^{V\frac{N_f}{2}L_0-V|p|L_I-V (N_f/2-|p|)L_0}, \hspace{2cm}
-N_f/2 \leq p \leq 0 \nn \\ 
& = & e^{V|p|(L_0-L_I)} = e^{V|p|L_B}.
\ee
Note that it smoothly connects to the $p\geq 0$ result (\ref{pgeq0}).

Combining the above results we find
\be
\langle e^{2ip\theta'}\rangle_{N_f} = e^{- V L_B ( |p+N_f/2|-N_f/2 )}
\label{any}
\ee
for any integer values of $p$.

The general result (\ref{any}) implies that the $\theta$-distribution is a
Lorentzian times the phase factor: 
\be\label{th-distmuGmpi2UQ}
\langle\delta(2\theta-2\theta')\rangle_{1+1}=
\frac{1}{\pi} \sum_{p=-\infty}^\infty \ e^{-2i \theta p}e^{- V L_B (|p+1|-1)}
= e^{2i\theta}
\frac{e^{V L_B}}{\pi} \frac{\sinh(V L_B)}{\cosh(V L_B)-\cos(2\theta)}.
\ee
As we have seen previously, the unquenched 1+1 distribution is related to the
phase quenched 1+1* distribution
\be
\langle\delta(2\theta-2\theta')\rangle_{1+1} 
= e^{2i\theta}\frac{Z_{1+1^*}}{Z_{1+1}}\langle
\delta(2\theta-2\theta')\rangle_{1+1^*}. 
\ee
Comparing this with Eq.~(\ref{thdist-Qmu>mpiO2v2}) and Eq.~(\ref{th-distmuGmpi2UQ}) we
see that the quenched and phase 
quenched $\theta$-distributions are identical also for $\mu > m_\pi/2$.

In conclusion, we have shown that the $\theta$-distribution is nonanalytic at
the point where the quark mass enters the support of the Dirac spectrum.  
This implies, for example, that the distribution of the phase in this regime
cannot be obtained by analytic continuation from imaginary values of $\mu$
(see \cite{SplSve,Conradi} for a discussion of the analytic continuation
of the phase factor to imaginary values of the chemical potential).

\section{The distribution of the Baryon number and the Chiral condensate
  ($\mu>m_\pi/2$)} 

In this section we compute the distribution of the baryon number and  the
chiral condensate over the phase angle. As for the distribution of the angle
itself we will work to leading order in chiral perturbation theory
which is the  mean field result for $\mu>m_\pi/2$. 

\subsection{The baryon number}

In order to work out $\langle n_B \delta(2\theta-2\theta')\rangle_{N_f}$ we need 
the moments 
\be
\frac{1}{Z_{N_f}} \langle
\frac{\det^{p}(D+\mu\gamma_0+m)}{\det^{p}(D-\mu\gamma_0+m)} \det(D+\tilde\mu\gamma_0+m)^{N_f} \rangle,
\ee
where the chemical potential for the $N_f$ quarks is denoted by
$\tilde\mu$. The distribution is then obtained after differentiation w.r.t.
$\tilde\mu$ at $\tilde\mu = \mu$, 
multiplication by $\exp(-2ip\theta)$ and summation over $p$.   

For $p\geq 0$ the bosonic mean field rules discussed in previous section
lead to  a factorization of the
moments  as follows
\be
\frac{1}{Z_{N_f}} \frac{\left \langle
\det^{2p}(D+\mu\gamma_0+m)\det(D+\tilde\mu\gamma_0+m)^{N_f} \right\rangle}
{\left\langle\det^{p}(D-\mu\gamma_0+m)\det^{p}(D+\mu\gamma_0+m)  \right \rangle}.
\ee
Since $|\tilde\mu-\mu|<m_\pi$ there is no condensation of pions for the
partition function in the
numerator. It follows that there is no dependence on $\tilde\mu$ and hence
all terms with $p\geq0$ vanish after differentiation w.r.t. $\tilde\mu$.

When $p$ is negative the $\det^{|p|}(D-\mu\gamma_0+m)$ is in the numerator
and condensation of pions occurs. This leads to a dependence on $\tilde\mu$
through the mean field Lagrangian
\be
L_I(-\mu,\tilde\mu)=-2F^2(\mu+\tilde\mu)^2/4-\frac{2\Sigma^2m^2}{(\mu+\tilde\mu)^2F^2}.
\ee    
Note that this reduces to $L_I$ given in (\ref{LI}) for $\tilde\mu=\mu$.

As in the previous section we must consider separately the cases 
$-N_f/2 \le p <0$
and $p<-N_f/2$. For $-N_f/2 \le p<0$ the moments are given by
\be
\frac{1}{Z_{N_f}} 
\langle e^{2pi\theta'(\mu)}{\det}^{N_f}(D+\tilde\mu\gamma_0+m)  \rangle
= e^{-2|p|V L_I(-\mu,\tilde\mu)+|p|V L_I(-\mu,\mu)+|p|V L_0 }
\ee
at mean field level. While for $p<-N_f/2$ we find
\be
\frac{1}{Z_{N_f}} 
\langle e^{2pi\theta'(\mu)}{\det}^{N_f}(D+\tilde\mu\gamma_0+m)  \rangle
=e^{-N_f V L_I(-\mu,\tilde\mu)+|p|V L_I(-\mu,\mu)-(|p|-N_f)V L_0}.
\ee
In both cases the derivative w.r.t. $\tilde\mu$ pulls down the prefactor $V
[d/d{\tilde\mu}] L_I(-\mu,\tilde\mu)$ but multiplied with a different 
numerical factor. This leads to
\be
&&\langle n_B \delta(2\theta-2\theta')\rangle_{N_f} 
=\frac{1}{\pi} \Big[\frac d{d {\tilde\mu}} L_I(-\mu,\tilde\mu)\Big]_{\tilde\mu=\mu}\\
&&\hspace{1cm}\times\left(\sum_{-N_f/2\leq p<0}(-2)|p|e^{-2ip\theta}e^{-V L_B(|p+N_f/2|-N_f/2)} 
     +\sum_{p<-N_f/2}(-N_f)e^{-2ip\theta}e^{-V L_B(|p+N_f/2|-N_f/2)}\right).\nn
\ee
For $N_f=2$ there is only one term in the first sum, namely $p=-1$, and it
can be included in the second sum
\be
&&\langle n_B \delta(2\theta-2\theta')\rangle_{1+1} 
=\frac{1}{\pi} \Big[\frac{VL_I}{\mu}\Big](-2)\sum_{p\leq -1}e^{-2ip\theta}e^{-V L_B(|p+1|-1)}.
\ee
The sum can be performed analytically,
\be
&&\langle n_B \delta(2\theta-2\theta')\rangle_{1+1} 
=-\frac{2}{\pi} \Big[\frac{VL_I}{\mu}\Big]
e^{2i\theta}e^{2VL_B} \frac{1}{e^{VL_B}-e^{2i\theta}}
=-\frac{2}{\pi} \Big[\frac{VL_I}{\mu}\Big]
e^{2VL_B}\frac {-1}{2i} \frac d{d\theta}\log( e^{VL_B}-e^{2i\theta}).
\ee
The total baryon number density is given by the integral over the
distribution (recall that $2\theta\in[-\pi,\pi]$) and vanishes. 
The distribution of the baryon number over the phase angle 
is proportional to a total derivative but not of
the distribution of the phase as was the case for $\mu < m_\pi/2$.

\subsection{The chiral condensate}

As in section \ref{sec:Sigma} we now dentoe the mass of the $N_f$ quarks by
$\tilde{m}$ and differentiate with respect to this mass. The computation is
somewhat analogous to the one given in the previous
section except that the terms with positive $p$ also contribute.
For $N_f=2$ we find
\be
\langle \bar{\psi}\psi \ \delta(2\theta-2\theta')\rangle_{1+1} 
= \frac{2\Sigma V}{\pi}\sum_{p=0}^\infty e^{-2ip\theta}e^{-V L_B(|p+1|-1)} 
-\frac{2}{\pi}\Big[\frac{d}{d\tilde{m}}L_I(m,\tilde{m})\Big]_{\tilde{m}=m}
\sum_{p=-\infty}^{-1}e^{-2ip\theta}e^{-V L_B(|p+1|-1)} 
\ee
where
\be
L_I(m,\tilde{m}) = -2\mu^2F^2-\frac{\Sigma^2(m+\tilde{m})^2}{8\mu^2F^2}. 
\ee
The sums can be rewritten as
\be
\langle \bar{\psi}\psi \ \delta(2\theta-2\theta')\rangle_{1+1} 
= \frac{2\Sigma V}{\pi}\frac{e^{-VL_B}}{e^{2i\theta}-e^{-VL_B}}
+\frac{2\Sigma V}{\pi}
+\frac{2}{\pi}\frac{\Sigma^2m}{\mu^2F^2}
e^{2i\theta}e^{VL_B} \frac{e^{VL_B}}{e^{VL_B}-e^{2i\theta}}.
\ee      
The first and the last term both vanish upon integration over 
$\theta \in [-\pi/2,\pi/2]$ and
 leaves, $2V\Sigma$, which, after dividing by the volume,
 is the expected mean field value of the chiral
condensate for $N_f =2$. 
Note that the amplitude of the first term is exponentially small while that
of the last term is exponentially big. The severe cancellations which take
place upon integration of the last term over $\theta$ are just like those for
the baryon number.

\section{QCD in one Euclidean dimension}
\label{sec:1dQCD}

In this section  we will show that for one dimensional QCD 
the distribution of the
phase of the fermion determinant
changes from Gaussian to Lorentzian shape when the quark mass enters the
Dirac spectrum.

Lattice QCD in one Euclidean dimension (time only) with gauge group $U(N_c)$ 
is sufficiently simple that we can solve the partition function and moments
of the phase factor analytically starting from the fundamental partition
function. The reason is twofold.  First,  there is no Yang-Mills action, and
second, the staggered Dirac operator, $M$,
can be reduced to the determinant a $N_c\times N_c$ matrix \cite{BD}
\be
\det M =  2^{-nN_c}\det[e^{n\mu_c} + e^{-n\mu_c}+ e^{n \mu} U +e^{-n \mu}
U^\dagger], 
\label{det1}
\ee
where $U\in U(N_c)$. The analogue of  $m_\pi/2$
 or $m_N/N_c$
 is $\mu_c =
\sinh^{-1} m$ and $n$ is the number of lattice points. The 
eigenvalues  of $M$ are located on an ellipse of width $\sinh\mu$ along the
real axis. This means that the quark mass is inside the eigenvalue domain when
$\mu>\mu_c$.
In the limit $n N_c \to \infty$ the ratio of the
full partition function and the phase quenched partition function
 approaches one for the $SU(N_c)$ theory whereas this ratio 
the $U(N_c)$ partition functions shows a phase transition when the 
quarks mass enters the Dirac spectrum exactly as in QCD \cite{RV}. 
For this reason
we study the $U(N_c)$ lattice model rather than the $SU(N_c)$ lattice model.
In adddtion,  the $U(N_c)$ model is mathematically simpler than the
$SU(N_c)$ lattice model. 
  The partition function is defined by 
\be
Z_{N_f}(\mu_c,\mu) = \int_{U(N_c)} dU \det M.
\label{eq8}
\ee
Despite its simplicity many interesting things can be learned from QCD in one
dimension. For example, in \cite{RV} it was found that spectral density of
the Dirac operator is a highly oscillatory function 
when the quark mass is inside the
ellipse of eigenvalues, and that the link between these oscillations and the
chiral condensate is exactly the same 
as what was found for 4d QCD with dynamical quarks \cite{OSV}.  

Below we will show that the distribution of the phase of the fermion
determinant in one dimensional QCD also undergoes a transition from a
Gaussian to a Lorentzian shape when the quark mass enters the
eigenvalue spectrum. For simplicity we only work out the quenched
distribution. As above we start from the moments of the phase factor 
\be
\langle e^{2ip\theta'} \rangle
&=& \int_{U(N_c)} dU \frac { {\det}^p M}{ {\det}^p M^\dagger}. 
\ee
Notice that the expectation value only depends on $|p|$.
Following \cite{RV}, where the first moment ($p=1$) was worked out, we rewrite
the $U$-integral as
\be
\langle e^{2ip\theta'} \rangle
&=& \int_{U(N_c)} dU \frac{{\det}^p (1-U e^{n\mu-n\mu_c})
{\det}^p(1- U^\dagger e^{-n\mu-n\mu_c})}
{{\det}^p (1-U e^{-n\mu-n\mu_c}){\det}^p(1- U^\dagger e^{n\mu-n\mu_c})}.
\label{mom-1d}
\ee
In this form the Conrey-Farmer-Zirnbauer formula \cite{CFZ} can be applied
directly for $\mu<\mu_c$. In the large $N_c$ limit the result simplifies to 
\be
\langle e^{2ip\theta'} \rangle = \langle e^{2i\theta'} \rangle^{p^2} = \left(1
  -  \frac {\mu^2}{\mu_c^2}\right)^{p^2}. 
\ee
If $\mu_c$ is interpreted as the chemical potential for which the quark mass
enters the eigenvalue domain, this is exactly the same
form as we obtained for the $\epsilon$-regime of QCD in section
\ref{sec:epsilon}. Hence we find the expected Gaussian form for the distribution
of the phase of the fermion determinant, 
\be
\langle\delta(\theta-\theta')\rangle = \frac{1}{\sqrt{\pi
    \Omega}}e^{-\theta^2/\Omega}
\label{th-dist-1d} \qquad {\rm for} \qquad    {\rm \mu<\mu_c, \  N_c\to \infty},
\ee
where $\Omega\equiv -\log(1-\mu^2/\mu_c^2)$.

For $\mu>\mu_c$ the conditions for applying the Conrey-Farmer-Zirnbauer
formula directly are violated. Now, however, we instead can rewrite the
determinants containing $U^\dagger$ as 
\be
\frac
{{\det}^p(1- U^\dagger e^{-n\mu-n\mu_c})}
{{\det}^p(1- U^\dagger e^{n\mu-n\mu_c})}
=
e^{-2pnN_c\mu}\frac{{\det}^p(1-  Ue^{n\mu+n\mu_c})}
{{\det}^p(1- U e^{-n\mu+n\mu_c})},
\label{rewrite}
\ee
so that the entire integrand in Eq.~(\ref{mom-1d}) only depends on $U$. This
implies that when we  expand the denominator in $U$ (which is allowed for 
$\mu>\mu_c$) only the constant term is nonzero upon integration over $U$. 
Using that the moments of the phase factor only depend on $|p|$
we obtain the exact result
\be
\langle e^{2ip\theta'} \rangle  =e^{-2n|p|N_c \mu}, \qquad
{\rm for} \quad \mu>\mu_c .
\label{evenmom}
\ee
Summing over  $p$ in Eq.~(\ref{evendist}) results in the compact Lorentzian
c.f. Eq.~(\ref{thdist-Qmu>mpiO2}) 
\be
\label{1dQCD-Lorentzian}
\langle\delta(2\theta-2\theta')\rangle 
= \frac{1}{\pi}\frac{\sinh(2nN_c \mu)}{\cosh(2nN_c \mu)-\cos(2\theta)} \qquad {\rm for} \qquad 
\mu>\mu_c \ , \ 2\theta \in[-\pi,\pi]. 
\ee
We stress that this exact result is valid for any value of $N_c$.  

\begin{figure}[t!]
\unitlength1.0cm
\epsfig{file=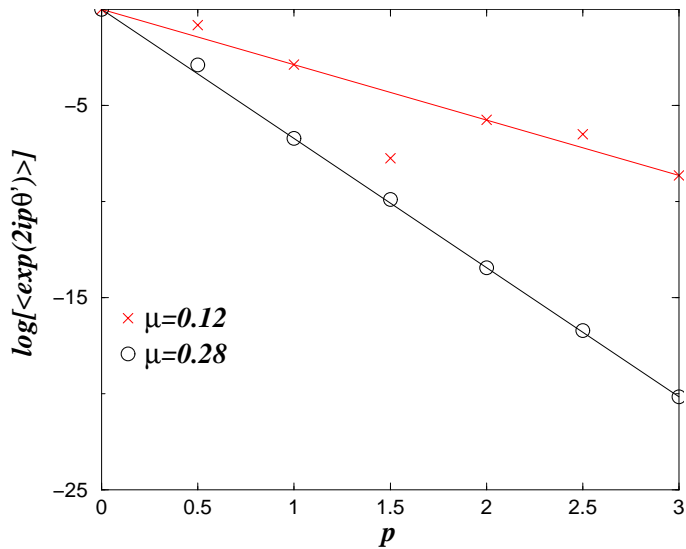,clip=,width=9cm}
\caption{\label{fig:replica}Numerical evaluation of the quenched moments of the
  average phase factor in one dimensional QCD versus $p$ for $\mu_c=0.1$, $n = 4$ and
  $N_c = 3$. As indicated by the lines the even moments join smoothly in
  accordance with (\ref{evenmom}). However, the even and odd moments are not
  smoothly connected for $0<\mu_c<\mu$.} 
\end{figure}

Note that we have computed the distribution of the phase angle of
the square of the fermion determinant, i.e. of $2\theta$. 
The reason is that this does not require the use of
the replica trick. 
By comparing the numerical result for half-integer moments with the analytical
result (\ref{evenmom})  obtained for integer moments one finds that the replica
trick does not work when quark mass is inside the eigenvalues. See figure
\ref{fig:replica}.  
The only exception is the case $\mu_c=0$: Then the rewriting in
Eq.~(\ref{rewrite}) results in $2p$ powers of the determinants 
\cite{Martin-pr} which are then well-defined for half-integer $p$.  The
expression for the odd moment when $\mu_c=0$ is therefore also given by
(\ref{evenmom}).

\newpage

\section{Distribution of $f=\log |\det(D+\mu\gamma_0+m)|/\det(D+m)$
  for $\mu<m_\pi/2$} 
\label{sec:Fidst}

So far we have considered distributions of the phase of the
fermion determinant. As we now show it is also possible to compute the
distributions as a function of the absolute value of the fermion
determinant. We will do this to one loop order in chiral perturbation theory
using the replica trick.  
Since only the case $\mu<m_\pi/2$ will be considered there are no issues with
the use of the replica trick.  

Since the absolute value of the fermion determinant depends on the large
eigenvalues of the Dirac operator we analyze the distribution of
$f\equiv \log[|\det(D+\mu\gamma_0+m)|/\det(D+m)]$ which, as we shall see below, 
depends only
on the finite difference of the one-loop free energy at $\mu$ and at $\mu=0$. 
In  \cite{Ejiri} the distribution of $F\equiv \exp(f)$ was studied in lattice QCD
using the Taylor expansion method. Since determinants fluctuate by many orders of
magnitude we feel that it is more appropriate to analyze the distribution
of the logarithm of their magnitude 
instead. The two distributions are related by a simple
transformation
\be
\langle \delta(f-f')\rangle = F \langle \delta(F-F')\rangle,
\ee
where $f'$ is the magnitude of the logarithm 
of the ratio of the determinants -- its fluctuations are induced by
the gauge field fluctuations.

To compute
the distribution of the magnitude of the logarithm of the determinants
we rewrite the 
$\delta$-function  as
\be
       \langle \delta(f-f')\rangle_{1+1} 
 & = & \int_{-\infty}^\infty \frac{dp}{2\pi} \langle 
 e^{-ip(f-f')} \rangle_{1+1}\nn \\
& = & \frac{1}{Z_{1+1}}\int_{-\infty}^\infty \frac{dp}{2\pi}
e^{-ipf}\left\langle\left(
\frac{\det(D+\mu\gamma_0+m)\det(D-\mu\gamma_0+m)}{{\det}^2(D+m)}
\right)^{\frac{ip}{2}}{\det}^2(D+\mu\gamma_0+m)\right\rangle.
\ee
For even $ip$  the average can be interpreted as a partition function
with bosonic and fermionic flavors. 
We will calculate this partition function
to one loop order in 
chiral perturbation theory. Since we consider the magnitude
of the determinant we expect that the moments will be analytic in $p$ and
can be analytically continued to imaginary $ip$
 As far as we know this is the first case where the replica trick is used
this way. 

Using the same one-loop combinatorics as before, we find after analytical
continuation to imaginary $ip$,
\be
\langle\delta(f-f')\rangle_{1+1} 
& = & \int_{-\infty}^\infty \frac{dp}{2\pi}e^{-ip(f-E_f) -\frac 12 \sigma_f^2 p^2},  
\ee
where
\be
E_f= 2\Delta G_0(\mu) - 4 \Delta G_0(\mu/2), \\
\sigma^2_f = \frac 12 \Delta G_0 (\mu) - 2 \Delta G_0(\mu/2). 
\ee
The integral over $p$ is Gaussian and can be evaluated by completing squares.
This results in
\be
\label{fdist1+1}
\langle \delta(f-f')\rangle_{1+1}  = \frac 1{\sigma_f \sqrt {2\pi}} 
e^{-\frac{(f-E_f)^2}{2\sigma_f^2}}.
\ee
Both $E_f$ and $ \sigma^2_f$ are positive. In the thermodynamic limit
at nonzero $T$ and $\mu$ we can see this using Eq.~(\ref{g0familiar})
\be
\sigma^2_ f
 & = & \frac{Vm_\pi^2T^2}{2\pi^2}\sum_{n=1}^\infty 
\frac{K_2(\frac{m_\pi n}{T})}{n^2}
\left[\cosh(\frac{2\mu n}{T}) -4 \cosh(\frac{\mu n}{T})+3\right], \nn \\
 & = & \frac{Vm_\pi^2T^2}{2\pi^2}\sum_{n=1}^\infty \frac{K_2(\frac{m_\pi n}{T})}{n^2}8\sinh^4(\frac{\mu n}{2T}).
\ee 
Similarly we can write $E_f$ as
\be
E_f&=&
\frac{Vm_\pi^2T^2}{\pi^2}\sum_{n=1}^\infty \frac{K_2(\frac{m_\pi n}{T})}{n^2}
\left[\cosh(\frac{2\mu n}{T}) -2\cosh(\frac{\mu n}{T})+1\right], \nn \\
 & = & 2\frac{Vm_\pi^2T^2}{\pi^2}\sum_{n=1}^\infty \frac{K_2(\frac{m_\pi n}{T})}{n^2}\cosh(\frac{\mu n}{T})(\cosh(\frac{\mu n}{T})-1).
\ee 
Exactly the same combinatorics can be applied to the finite $L$ expressions
for $\sigma^2_f$ and $E_f$ (see (\ref{g0finiteL}))
 resulting in the positivity of these quantities
at finite $L$ and $L_0$.

\begin{figure}[t!]
\unitlength1.0cm
\epsfig{file=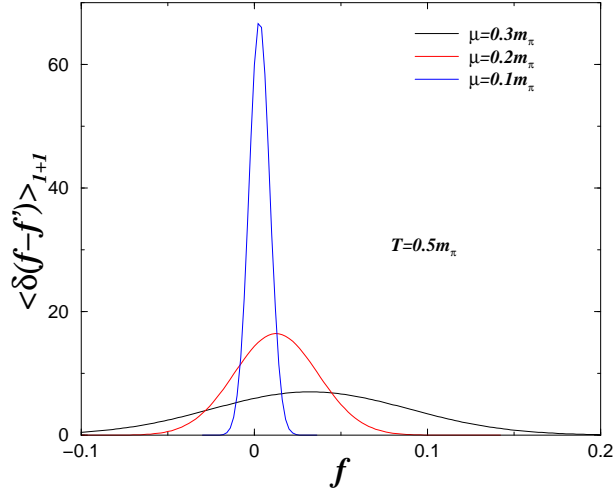,clip=,width=8cm}
  \caption{\label{fig:F-dist}{\bf The 
      $f$-distribution in 1-loop chiral perturbation theory.}   
     The distribution of the partition function with
    $f=\log(|\det(D+\mu\gamma_0+m)|/\det(D+m))$ in a box with
    $Vm_\pi^4=10$. The temperature is fixed and as $\mu$ increases the
    distribution becomes broader and moves away from zero.}
\end{figure}

Let us make a simple cross check of the formula for $E_f$ and $\sigma^2_f$. 
Since $f=0$ for $\mu=0$
the expectation value and the variance of the $f$-distribution must vanish
in the limit $\mu \to 0$ which is indeed the case
(see figure \ref{fig:F-dist}).

In oder to better understand the structure of the result 
it is useful to work out the
combinatorics for an arbitrary number of flavors $N_f$
\be
\langle\delta(f-f')\rangle_{N_f} = \frac 1{\sigma_f \sqrt {2\pi}} 
e^{-\frac{(f-N_fE_f/2)^2}{2\sigma_f^2}}.
\ee 
We note that, an increasing number of flavors simply shifts the average value
of $f$.

\subsection{The distribution of the baryon number over $f$}
\label{subsec:nBdistF}

Even though the baryon number is zero when evaluated in chiral
perturbation theory it does not necessarily vanish  when evaluated
for a constrained fermion determinant. In section \ref{sec:baryon}
we derived the distribution of the baryon number for fixed  phase. Here we
compute the distribution of the baryon number as a function of $f$. 

In order to compute $\langle n_B \delta(f-f')\rangle_{1+1}$ we denote the
chemical potential in the  usual two flavor determinant by
$\tilde{\mu}$ instead of $\mu$, then differentiate with respect to 
$\tilde{\mu}$ and finally take the limit
$\tilde{\mu}\to \mu$. The 
$\delta$ function is represented as in the previous sections
\be
&& \langle n_B \ \delta(f-f')\rangle_{1+1} \\
 & = & \frac{1}{2Z_{1+1}}\lim_{\tilde{\mu}\to\mu}\frac{d}{d\tilde{\mu}}\int_{-\infty}^\infty \frac{dp}{2\pi}
e^{-ipf}\left\langle\left(
\frac{\det(D+\mu\gamma_0+m)\det(D-\mu\gamma_0+m)}{{\det}^2(D+m)}
\right)^{\frac{ip}{2}}{\det}^2(D+\tilde{\mu}\gamma_0+m)\right\rangle.\nn
\ee
To one loop order in chiral perturbation theory this becomes
\be
& & \langle n_B \ \delta(f-f')\rangle_{1+1} =
 \big[\frac{d}{d\tilde{\mu}}\left(G_0(-\mu,\tilde{\mu})-2G_0(0,\tilde{\mu})\right)\big]_{\tilde{\mu}=\mu} \int_{-\infty}^\infty \frac{dp}{2\pi} \ ip \
e^{-ip(f-E_f) -\frac 12 \sigma_f^2 p^2}.
\ee
The Gaussian integral over $p$ results in
\be
& & \langle n_B \ \delta(f-f')\rangle_{1+1} =
- \big[\frac{d}{d\tilde{\mu}}\left(G_0(-\mu,\tilde{\mu})-2G_0(0,\tilde{\mu})\right)\big]_{\tilde{\mu}=\mu} \frac{E_f-f}{\sigma_f^2}
\langle\delta(f-f')\rangle_{1+1} . 
\ee
The baryon number operator is not positive definite and neither is its
distribution over $f$. It changes sign at the expectation value of
the Gaussian distribution so that the total baryon density vanishes
\be
  \langle n_B \rangle_{1+1}=\int_{-\infty}^\infty df\ \langle n_B \
  \delta(f-f')\rangle_{1+1} = 0.
\ee
As is the case 
for the distribution of the baryon number over $\theta$ the zero
value can also be obtained by noting that the integrand is a total
derivative.

\subsection{The distribution of the chiral condensate over $f$}
\label{subsec:SigmadistF}

In this section we derive the distribution of the chiral condensate over $f$. As above
we represent $\delta(f-f')$ by an integral over the moments so that 
\be
&& \langle \bar{\psi}\psi \ \delta(f-f')\rangle_{1+1} \nn\\
& = & 
\frac{1}{Z_{1+1}}\lim_{\tilde{m}\to m}\frac{d}{d\tilde{m}}\int_{-\infty}^\infty \frac{dp}{2\pi}
e^{-ipf}\left\langle\left(
\frac{\det(D+\mu\gamma_0+m)\det(D-\mu\gamma_0+m)}{{\det}^2(D+m)}
\right)^{\frac{ip}{2}}{\det}^2(D+\mu\gamma_0+\tilde{m})\right\rangle.
\ee
The combinatorics of  possible one-loop contributions of Goldstone bosons 
leads to
\be
&&\langle \bar{\psi}\psi \ \delta(f-f')\rangle_{1+1} \\
& = & 
\int_{-\infty}^\infty \frac{dp}{2\pi}   \big[ip
\frac{d}{d\tilde{m}}\left(G_0(\mu,\mu,m,\tilde{m})+G_0(-\mu,\mu,m,\tilde{m})-2G_0(0,\mu,m,\tilde{m})\right)\nn\\
&& \hspace{2cm} +4\frac{d}{d\tilde{m}}\left(G_0(0,\tilde{m})-G_0(0,m)
  \right)\big]_{\tilde{m}=m}   
e^{ -\frac 12 \sigma_f^2 p^2 +ip(E_f-f)},\nn \\ 
&=& \langle \bar \psi \psi \rangle_{1+1} \langle \delta(f-f')\rangle_{1+1}
- 
\frac{d}{d\tilde{m}}\big[G_0(\mu,\mu,m,\tilde{m})+G_0(-\mu,\mu,m,\tilde{m})-2G_0(0,\mu,m,\tilde{m})\big]_{\tilde{m}=m}
\frac{E_f-f}{\sigma_f^2}
\langle\delta(f-f')\rangle_{1+1} . \nn
\ee  
The first term in the last line 
gives the chiral condensate upon integration
over $f$ while the second term in the final line integrates to zero
in precisely the same way as in the case of the baryon density. 

\subsection{The  $f$-distribution evaluated in an ensemble generated at $\mu=0$}
\label{subsec:FdistrEjiri}

In \cite{Ejiri} the distribution of 
$F=|\det(D+\mu\gamma_0+m)|/\det(D+m)$
is studied in lattice QCD for  an ensemble generated at $\mu=0$. We will
again study the distribution of  $f\equiv \log F$ for this case. It is given by
\be
 &   & \frac{1}{Z_{1+1}(\mu=0)}\langle\delta(f-f'){\det}^2(D+m)\rangle \\
 & = & \frac{1}{Z_{1+1}(\mu=0)}\int_{-\infty}^\infty \frac{dp}{2\pi}
e^{-ipf}\left\langle\left(
\frac{\det(D+\mu\gamma_0+m)\det(D-\mu\gamma_0+m)}{{\det}^2(D+m)}
\right)^{\frac{ip}{2}}{\det}^2(D+m)\right\rangle.\nn
\ee
When evaluated to one loop order in chiral perturbation theory we find
\be
\frac{1}{Z_{1+1}}\langle\delta(f-f'){\det}^2(D+m)\rangle 
= \int_{-\infty}^\infty \frac{dp}{2\pi}
e^{-ip(f-\tilde E_f) -\frac 12 \sigma_f^2p^2}  =   \frac 1{\sigma_f \sqrt{2\pi}}e^{-(f-\tilde E_f)^2/(2\sigma_f^2)},
\ee
where
\be
\tilde E_f = 2\Delta G_0(\mu/2).
\ee
In comparison to (\ref{fdist1+1}) we see that only the expectation value of
$f$ has changed whereas the variance takes the same value as in previous
sections. 

\section{Constraining both $\theta$ and $F$ for  ($\mu<m_\pi/2$)}

In oder to understand what happens if both the phase and the
magnitude of the fermion determinant are fixed we need to compute the
correlation between the moments of the phase factor and $F$.

Let us consider the correlation of any moment of the phase factor and $F$
\be
&&\hspace{-3cm}\left\langle
\frac{{\det}^p(D+\mu\gamma_0+m)}{{\det}^p(D-\mu\gamma_0+m)}
\frac{\det^q(D+\mu\gamma_0+m)\det^q(D-\mu\gamma_0+m)}{\det^q(D+m)\det^q(D+m)}
\right\rangle \nn\\
&& - \left\langle\frac{{\det}^p(D+\mu\gamma_0+m)}{{\det}^p(D-\mu\gamma_0+m)}
\right\rangle\left\langle\frac{\det^q(D+\mu\gamma_0+m)\det^q(D-\mu\gamma_0+m)}
{\det^q(D+m)\det^q(D+m)}\right\rangle
\nn\\
&& \hspace{-3.4cm} = \ e^{-p^2\Delta G_0+q^2\Delta G_0-4q^2\Delta G_0(\mu/2)} -e^{-p^2\Delta G_0}e^{q^2\Delta G_0-4q^2\Delta G_0(\mu/2)} = 0.
\ee
The reason is that terms linear in $p$ in the first exponent cancel
completely. In other words, even though there are bound states 
(Goldstone bosons) with
non zero charge which potentially can couple the phase factor to the
absolute value of the determinant, their contributions exactly cancel
each other. This is also the case if the average is calculated for 
$N_f$ dynamical flavors.         

We have thus shown there are no correlations between the absolute value
of the fermion determinant and the phase to one-loop order in chiral 
perturbation
theory (for $\mu<m_\pi/2$). Hence we automatically find
\be
  \langle \delta(f-f')\delta(\theta-\theta')\rangle_{1+1}
= \langle \delta(f-f')\rangle_{1+1} \langle \delta(\theta-\theta')\rangle_{1+1}
\ee
and 
\be
  \langle n_B\delta(f-f')\delta(\theta-\theta')\rangle_{1+1}
= \langle n_B\delta(f-f')\rangle_{1+1} \langle
\delta(\theta-\theta')\rangle_{1+1}+\langle \delta(f-f')\rangle_{1+1} \langle
n_B\delta(\theta-\theta')\rangle_{1+1} .
\ee
One can convince oneself that this factorization does not hold 
for $\mu>m_\pi/2$.

\section{Conclusions}
\label{sec:concl}

The distribution of the phase of the fermion determinant for QCD with nonzero
quark chemical potential has been computed to leading order 
in chiral perturbation
theory. When the quark mass is outside the support of the 
Dirac spectrum (small $\mu$) the distribution becomes Gaussian whereas the
distribution is Lorentzian (modulo $2\pi$) when the quark mass is inside the
support. This non-analytic behavior is also found for QCD in one Euclidean
dimension by a direct evaluation of the involved partition functions.

The distribution of the baryon number and the chiral condensate as a function
over the phase angle has also been computed in chiral perturbation theory. The
results show analytically that extreme cancellations are essential for the 
vacuum expectation values of these fundamental quantities.   

The ratio of the magnitude of the fermion determinant to its value at $\mu=0$
is ultraviolet finite and can be studied within chiral perturbation
theory. We have computed the distribution of the logarithm of this ratio, 
$f$, as well as the
distribution of the baryon number and the chiral condensate over
$f$. Contrary to the $\theta$-distribution the distribution of $f$ is real
and positive. In fact, within one-loop chiral perturbation theory for
$\mu<m_\pi/2$ there are no correlations between the phase and the absolute
value of the fermion determinant.   

The results obtained here are complementary  to lattice results obtained by
Ejiri \cite{Ejiri}. The results for one-loop chiral perturbation theory
when the quark mass is outside the eigenvalue distribution of the Dirac
operator, confirms the Gaussian shape of the $\theta$-distribution first found
in lattice simulations \cite{Ejiri}. The
analytical results, however, also show that exponentially large cancellations
may take place when integrating over $\theta$. Not only are these cancellations
essential in order to measure the baryon number and the
chiral condensate correctly, the extreme tail of the distribution may 
contribute
significantly to the final result. A small non Gaussian term in the tail
of the $\theta$-distribution therefore could be the dominant term after
integration over $\theta$. The precise form of this tail is of course 
difficult to access numerically.  

The Lorentzian shape of the distribution of the phase valid for larger
values of the chemical potentially shows that one should not take for granted
that the conditions for
the central limit theorem are satisfied. The nonanalyticity means that the
Lorentzian shape cannot be obtained by 
analytic continuation from imaginary values of the chemical potential. 
Since the Lorentzian form is present also for quenched QCD this prediction
can be tested in lattice QCD without worrying about the sign problem. Numerical
convergence is however expected to slow because of the large fluctuations of
the phase. If staggered fermions are used one also has to address the
issues raised in \cite{GSS}.       

Finally let us stress that both the Gaussian and the Lorentzian forms for the
$\theta$-distributions found here are leading order predictions of chiral
perturbation theory. It would be of considerable interest to work out the
next to leading order corrections. It seems natural that these terms
will give corrections to both 
the shape of the $\theta$-distribution and to its width.

The analytical work of this paper 
was inspired by  new developments in the numerical
density of states method. Such interplay between numerical lattice QCD and
analytical methods, 
is essential for progress towards our
understanding of strongly interacting
matter. In this paper this was illustrated by simulations of
one-dimensional lattice QCD.
Even if the analytical results do not yet offer 
a direct solution of the
sign problem, they allow us to better understand the regions where current
numerical methods can be applied \cite{splitrev}.

\vspace{4mm}

\noindent
{\sl Acknowledgments.} 
We wish to thank everybody at ``Tools for Finite Density QCD'' in Bielefeld
where part of this work was presented. Also thanks to Martin Zirnbauer, 
Dennis Dietrich, Gernot
Akemann and Poul Henrik Damgaard for discussions. MPL thanks the theoretical
high energy physics group at the University of Bielefeld for its hospitality.  
This work was supported  by U.S. DOE Grant No. DE-FG-88ER40388 (JV), the 
Danish Natural Science Research Council (KS).


\begin{thebibliography}{9}

\bibitem{endpoint}
M.~A.~Stephanov,
  %``Non-Gaussian fluctuations near the QCD critical point,''
  arXiv:0809.3450 [hep-ph];
  %%CITATION = ARXIV:0809.3450;%%
  %``QCD critical point and complex chemical potential singularities,''
  Phys.\ Rev.\  D {\bf 73}, 094508 (2006)
  [arXiv:hep-lat/0603014];
  %%CITATION = PHRVA,D73,094508;%%
%``QCD phase diagram and the critical point,''
  Prog.\ Theor.\ Phys.\ Suppl.\  {\bf 153}, 139 (2004)
  [Int.\ J.\ Mod.\ Phys.\  A {\bf 20}, 4387 (2005)]
  [arXiv:hep-ph/0402115].
  %%CITATION = IMPAE,A20,4387;%%


\bibitem{Nature}
  Y.~Aoki, G.~Endrodi, Z.~Fodor, S.~D.~Katz and K.~K.~Szabo,
  %``The order of the quantum chromodynamics transition predicted by the
  %standard model of particle physics,''
  Nature {\bf 443}, 675 (2006)
  [arXiv:hep-lat/0611014].
  %%CITATION = NATUA,443,675;%%

\bibitem{deFPh}
  P.~de Forcrand and O.~Philipsen,
  %``The curvature of the critical surface (m_ud,m_s)^{crit}(mu): a progress
  %report,''
  PoS {\bf LATTICE2008}, 208 (2008)
  [arXiv:0811.3858 [hep-lat]];
  %%CITATION = POSCI,LATTICE2008,208;%%
%\bibitem{deForcrand:2008vr}
  %  P.~de Forcrand and O.~Philipsen,
  %``The chiral critical point of Nf=3 QCD at finite density to the order
  %(mu/T)^4,''
  JHEP {\bf 0811}, 012 (2008)
  [arXiv:0808.1096 [hep-lat]].
  %%CITATION = JHEPA,0811,012;%%



\bibitem{owe1}
  P.~de Forcrand and O.~Philipsen,
 % ``The QCD phase diagram for small densities from imaginary chemical
  %potential,''
  Nucl.\ Phys.\ B {\bf 642}, 290 (2002)
  [arXiv:hep-lat/0205016].
  %%CITATION = HEP-LAT 0205016;%%


\bibitem{maria}
  M.~D'Elia and M.~P.~Lombardo,
  %``Finite density QCD via imaginary chemical potential,''
  Phys.\ Rev.\ D {\bf 67}, 014505 (2003)
  [arXiv:hep-lat/0209146].
  %%CITATION = HEP-LAT 0209146;%%

\bibitem{owe2}
  P.~de Forcrand and O.~Philipsen,
  %``The QCD phase diagram for three degenerate flavors and small baryon
  %density,''
  Nucl.\ Phys.\ B {\bf 673}, 170 (2003)
  [arXiv:hep-lat/0307020].
  %%CITATION = HEP-LAT 0307020;%%


\bibitem{ERL}
  M.~D'Elia, F.~Di Renzo and M.~P.~Lombardo,
  %``The strongly interacting Quark Gluon Plasma, and the critical behaviour
  %of QCD at imaginary chemical potential,''
  Phys.\ Rev.\  D {\bf 76}, 114509 (2007)
  [arXiv:0705.3814 [hep-lat]].
  %%CITATION = PHRVA,D76,114509;%%






\bibitem{BJ}
B.-J. Schäfer, {\sl The QCD phase structure from chiral effective models},
talk at the workshop ``Tools for finite density QCD''  Bielefeld
University. Slides available from  
http://www.physik.uni-bielefeld.de/igs/schools/Tools2008/tools.html.



\bibitem{gupta}
  R.~V.~Gavai and S.~Gupta,
%  ``Pressure and non-linear susceptibilities in QCD at finite chemical
  %potentials,''
  Phys.\ Rev.\ D {\bf 68}, 034506 (2003)
  [arXiv:hep-lat/0303013].
  %%CITATION = HEP-LAT 0303013;%%


\bibitem{Allton1}
  C.~R.~Allton {\it et al.},
%  ``The QCD thermal phase transition in the presence of a small chemical
  %potential,''
  Phys.\ Rev.\ D {\bf 66}, 074507 (2002)
  [arXiv:hep-lat/0204010].
  %%CITATION = HEP-LAT 0204010;%%


\bibitem{Allton2}
  C.~R.~Allton, S.~Ejiri, S.~J.~Hands, O.~Kaczmarek, F.~Karsch, E.~Laermann and C.~Schmidt,
  %``The equation of state for two flavor QCD at non-zero chemical  potential,''
  Phys.\ Rev.\ D {\bf 68}, 014507 (2003)
  [arXiv:hep-lat/0305007].
  %%CITATION = HEP-LAT 0305007;%%

\bibitem{Allton3}
  C.~R.~Allton {\it et al.},
 % ``Thermodynamics of two flavor QCD to sixth order in quark chemical
  %potential,''
  Phys.\ Rev.\ D {\bf 71}, 054508 (2005)
  [arXiv:hep-lat/0501030].
  %%CITATION = HEP-LAT 0501030;%%


%\cite{Endrodi:2009sd}
\bibitem{Endrodi:2009sd}
  G.~Endrodi, Z.~Fodor, S.~D.~Katz and K.~K.~Szabo,
  %``The curvature of the QCD phase transition line,''
  PoS {\bf LATTICE2008}, 205 (2008)
  [arXiv:0901.3018 [hep-lat]].
  %%CITATION = POSCI,LATTICE2008,205;%%


\bibitem{deFSW}
  P.~de Forcrand, M.~A.~Stephanov and U.~Wenger,
  %``On the phase diagram of QCD at finite isospin density,''
  PoS {\bf LAT2007}, 237 (2007)
  [arXiv:0711.0023 [hep-lat]].
  %%CITATION = POSCI,LAT2007,237;%%



\bibitem{Glasgow} 
  I.~M.~Barbour, S.~E.~Morrison, E.~G.~Klepfish, J.~B.~Kogut and
  M.~P.~Lombardo, 
  %``Results on finite density QCD,''
  Nucl.\ Phys.\ Proc.\ Suppl.\  {\bf 60} A (1998) 220.
  %[arXiv:hep-lat/9705042].
  %%CITATION = HEP-LAT 9705042;%%


\bibitem{fodor1}
  Z.~Fodor and S.~D.~Katz,
  %``Lattice determination of the critical point of QCD at finite T and mu,''
  JHEP {\bf 0203}, 014 (2002)
  [arXiv:hep-lat/0106002].
  %%CITATION = HEP-LAT 0106002;%%

\bibitem{fodor2}
  Z.~Fodor and S.~D.~Katz,
 % ``Critical point of QCD at finite T and mu, lattice results for physical
  %quark masses,''
  JHEP {\bf 0404}, 050 (2004)
  [arXiv:hep-lat/0402006].
  %%CITATION = HEP-LAT 0402006;%%


\bibitem{KW}
  F.~Karsch and H.~W.~Wyld,
  %``Complex Langevin Simulation Of The SU(3) Spin Model With Nonzero Chemical
  %Potential,''
  Phys.\ Rev.\ Lett.\  {\bf 55}, 2242 (1985).
  %%CITATION = PRLTA,55,2242;%%


\bibitem{FOC}
  J.~Flower, S.~W.~Otto and S.~Callahan,
  %``COMPLEX LANGEVIN EQUATIONS AND LATTICE GAUGE THEORY,''
  Phys.\ Rev.\  D {\bf 34}, 598 (1986).
  %%CITATION = PHRVA,D34,598;%%


\bibitem{AFP}
  J.~Ambjorn, M.~Flensburg and C.~Peterson,
  %``THE COMPLEX LANGEVIN EQUATION AND MONTE CARLO SIMULATIONS OF ACTIONS WITH
  %STATIC CHARGES,''
  Nucl.\ Phys.\  B {\bf 275}, 375 (1986).
  %%CITATION = NUPHA,B275,375;%%


\bibitem{aarts}
  G.~Aarts and I.~O.~Stamatescu,
  %``Stochastic quantization at finite chemical potential,''
  JHEP {\bf 0809}, 018 (2008)
  [arXiv:0807.1597 [hep-lat]];
  %%CITATION = JHEPA,0809,018;%%
  G.~Aarts,
  %``Can stochastic quantization evade the sign problem? -- the relativistic
  %Bose gas at finite chemical potential,''
  arXiv:0810.2089 [hep-lat]; 
  %G.~Aarts,
  %``Complex Langevin dynamics at finite chemical potential: mean field
  %analysis in the relativistic Bose gas,''
  arXiv:0902.4686 [hep-lat].
  %%CITATION = ARXIV:0902.4686;%%
  %%CITATION = ARXIV:0810.2089;%%


\bibitem{exp2ith-letter}
  K.~Splittorff and J.~J.~M.~Verbaarschot,
  %``Phase of the fermion determinant at nonzero chemical potential,''
  Phys.\ Rev.\ Lett.\  {\bf 98}, 031601 (2007)
  [arXiv:hep-lat/0609076].
  %%CITATION = PRLTA,98,031601;%%


\bibitem{phase-long}
  K.~Splittorff and J.~J.~M.~Verbaarschot,
  %``The QCD sign problem for small chemical potential,''
  Phys.\ Rev.\  D {\bf 75}, 116003 (2007)
  [arXiv:hep-lat/0702011].
  %%CITATION = PHRVA,D75,116003;%%


\bibitem{1loop}
  K.~Splittorff and J.~J.~M.~Verbaarschot,
  %``The Approach to the Thermodynamic Limit in Lattice QCD at \mu\neq0,''
  Phys.\ Rev.\  D {\bf 77}, 014514 (2008)
  [arXiv:0709.2218 [hep-lat]].
  %%CITATION = PHRVA,D77,014514;%%

\bibitem{Han}
  J.~Han and M.~A.~Stephanov,
  %``A Random Matrix Study of the QCD Sign Problem,''
  Phys.\ Rev.\  D {\bf 78}, 054507 (2008)
  [arXiv:0805.1939 [hep-lat]].
  %%CITATION = PHRVA,D78,054507;%%


\bibitem{BW}
  J.~C.~R.~Bloch and T.~Wettig,
  %``Random matrix analysis of the QCD sign problem for general topology,''
  arXiv:0812.0324 [hep-lat].
  %%CITATION = ARXIV:0812.0324;%%


\bibitem{OSV}
  J.~C.~Osborn, K.~Splittorff and J.~J.~M.~Verbaarschot,
  %``Chiral symmetry breaking and the Dirac spectrum at nonzero chemical
  %potential,''
  Phys.\ Rev.\ Lett.\  {\bf 94}, 202001 (2005);
  %[arXiv:hep-th/0501210].
  %%CITATION = HEP-TH 0501210;%%
  %J.~C.~Osborn, K.~Splittorff and J.~J.~M.~Verbaarschot, 
  %``Chiral Condensate at Nonzero Chemical Potential in the Microscopic Limit
  %of QCD,''
  arXiv:0805.1303 [hep-th].
  %%CITATION = ARXIV:0805.1303;%%



\bibitem{Gocksch}
  A.~Gocksch,
  %``SIMULATING LATTICE QCD AT FINITE DENSITY,''
  Phys.\ Rev.\ Lett.\  {\bf 61}, 2054 (1988).
  %%CITATION = PRLTA,61,2054;%%





\bibitem{Azcoiti}
  V.~Azcoiti, G.~Di Carlo, A.~Galante and V.~Laliena,
  %``New proposal for numerical simulations of theta-vacuum like systems,''
  Phys.\ Rev.\ Lett.\  {\bf 89}, 141601 (2002)
  [arXiv:hep-lat/0203017].
  %%CITATION = PRLTA,89,141601;%%

\bibitem{AN}
  K.~N.~Anagnostopoulos and J.~Nishimura,
  %``New approach to the complex-action problem and its application to a
  %nonperturbative study of superstring theory,''
  Phys.\ Rev.\  D {\bf 66}, 106008 (2002)
  [arXiv:hep-th/0108041].
  %%CITATION = PHRVA,D66,106008;%%

\bibitem{AANV}
 J.~Ambjorn, K.~N.~Anagnostopoulos, J.~Nishimura and J.~J.~M.~Verbaarschot,
  %``The factorization method for systems with a complex action: A test in
  %Random Matrix Theory for finite density QCD,''
  JHEP {\bf 0210}, 062 (2002)
  [arXiv:hep-lat/0208025].
  %%CITATION = HEP-LAT 0208025;%%

\bibitem{Schmidt}
  Z.~Fodor, S.~D.~Katz and C.~Schmidt,
  %``The density of states method at non-zero chemical potential,''
  JHEP {\bf 0703}, 121 (2007)
  [arXiv:hep-lat/0701022].
  %%CITATION = JHEPA,0703,121;%%

\bibitem{Ejiri}
  S.~Ejiri,
  %``On the existence of the critical point in finite density lattice QCD,''
  Phys.\ Rev.\  D {\bf 77}, 014508 (2008)
  [arXiv:0706.3549 [hep-lat]].
  %%CITATION = PHRVA,D77,014508;%%
  %arXiv:0706.3549 [hep-lat].
  %%CITATION = ARXIV:0706.3549;%%

\bibitem{misha}M. Stephanov, Phys. Rev. Lett. {\bf 76}, 4472 (1996).

\bibitem{SplitVerb2} K.~Splittorff and J.~J.~M.~Verbaarschot,
  %``Factorization of correlation functions and the replica limit of the  Toda
  %lattice equation,''
  Nucl.\ Phys.\ B {\bf 683}, 467 (2004).
  %[arXiv:hep-th/0310271].
  %%CITATION = HEP-TH 0310271;%%

\bibitem{AOSV}
G.~Akemann, J.~C.~Osborn, K.~Splittorff and J.~J.~M.~Verbaarschot,
%``Unquenched QCD Dirac operator spectra at nonzero baryon chemical
%potential,''
 Nucl.\ Phys.\ B {\bf 712}, 287 (2005).



\bibitem{SpecPhase}
  J.~C.~Osborn, K.~Splittorff and J.~J.~M.~Verbaarschot,
  %``Phase Diagram of the Dirac Spectrum at Nonzero Chemical Potential,''
  Phys.\ Rev.\  D {\bf 78}, 105006 (2008)
  [arXiv:0807.4584 [hep-lat]].
  %%CITATION = PHRVA,D78,105006;%%



\bibitem{STV}
K. Splittorff, D. Toublan, and J.J.M. Verbaarschot,
  Nucl.~Phys. B {\bf 620}, 290 (2002); %%CITATION = HEP-PH 0108040;%% 
 Nucl. Phys. B {\bf 639}, 524 (2002). %%CITATION = HEP-PH 0204076;%% 

\bibitem{CPT}
  J.~Gasser and H.~Leutwyler,
  %``Chiral Perturbation Theory To One Loop,''
  Annals Phys.\  {\bf 158}, 142 (1984);
  %%CITATION = APNYA,158,142;%%
% J.~Gasser and H.~Leutwyler,
  %``Chiral Perturbation Theory: Expansions In The Mass Of The Strange Quark,''
  Nucl.\ Phys.\  B {\bf 250}, 465 (1985).
  %%CITATION = NUPHA,B250,465;%%

\bibitem{HL}
  P.~Hasenfratz and H.~Leutwyler,
  %``Goldstone Boson Related Finite Size Effects In Field Theory And Critical
  %Phenomena With O(N) Symmetry,''
  Nucl.\ Phys.\ B {\bf 343}, 241 (1990).
  %%CITATION = NUPHA,B343,241;%%



\bibitem{AKW}
M.~Alford, A.~Kapustin, and F.~Wilczek,  Phys.~Rev. {\bf D 59} (1999) 054502. 
% [arXiv:hep-lat/9807039].

\bibitem{Edwards} 
S. F. Edwards and P.W. Anderson, J. Phys. {\bf F 5}, 965 (1975). 

\bibitem{replica}
  P.~H.~Damgaard and K.~Splittorff,
  %``Partially quenched chiral perturbation theory and the replica method,''
  Phys.\ Rev.\  D {\bf 62}, 054509 (2000)
  [arXiv:hep-lat/0003017].
 %%CITATION = PHRVA,D62,054509;%%


\bibitem{VS}
  E.~V.~Shuryak and J.~J.~M.~Verbaarschot,
  %``Random Matrix Theory And Spectral Sum Rules For The Dirac Operator In
  %QCD,''
  Nucl.\ Phys.\  A {\bf 560}, 306 (1993)
  [arXiv:hep-th/9212088].
  %%CITATION = NUPHA,A560,306;%%

\bibitem{V}
  J.~J.~M.~Verbaarschot,
  %``The Spectrum of the QCD Dirac operator and chiral random matrix theory: The
  %Threefold way,''
  Phys.\ Rev.\ Lett.\  {\bf 72}, 2531 (1994)
  [arXiv:hep-th/9401059].
  %%CITATION = PRLTA,72,2531;%%

\bibitem{DF}
  P.~H.~Damgaard and H.~Fukaya,
  %``The Chiral Condensate in a Finite Volume,''
  JHEP {\bf 0901}, 052 (2009)
  [arXiv:0812.2797 [hep-lat]].
  %%CITATION = JHEPA,0901,052;%%



\bibitem{SVZ}
  K.~Splittorff, J.~J.~M.~Verbaarschot and M.~R.~Zirnbauer,
  %``Nonhermitian Supersymmetric Partition Functions: the case of one bosonic
  %flavor,''
  Nucl.\ Phys.\  B {\bf 803}, 381 (2008)
  [arXiv:0802.2660 [hep-th]].
  %%CITATION = NUPHA,B803,381;%%


\bibitem{SV-bosonic}
  K.~Splittorff and J.~J.~M.~Verbaarschot,
  %``QCD with bosonic quarks at nonzero chemical potential,''
  Nucl.\ Phys.\  B {\bf 757}, 259 (2006)
  [arXiv:hep-th/0605143].
  %%CITATION = NUPHA,B757,259;%%


\bibitem{AP}
  G.~Akemann and A.~Pottier,
  %``Ratios of characteristic polynomials in complex matrix models,''
  J.\ Phys.\ A  {\bf 37}, L453 (2004)
  [arXiv:math-ph/0404068].
  %%CITATION = JPAGB,A37,L453;%%


\bibitem{Bergere}
  M.~C.~Bergere,
  %``Biorthogonal Polynomials for Potentials of two Variables and External
  %Sources at the Denominator,''
  arXiv:hep-th/0404126.
  %%CITATION = HEP-TH/0404

\bibitem{KSTVZ}
  J.~B.~Kogut, M.~A.~Stephanov, D.~Toublan, J.~J.~M.~Verbaarschot and
  A.~Zhitnitsky, 
  %``QCD-like theories at finite baryon density,''
  Nucl.\ Phys.\  B {\bf 582}, 477 (2000)
  [arXiv:hep-ph/0001171].
  %%CITATION = NUPHA,B582,477;%%

\bibitem{SS}
  D.~T.~Son and M.~A.~Stephanov,
  %``QCD at finite isospin density,''
  Phys.\ Rev.\ Lett.\  {\bf 86}, 592 (2001)
  [arXiv:hep-ph/0005225].
  %%CITATION = PRLTA,86,592;%%

\bibitem{SplSve}
  K.~Splittorff and B.~Svetitsky,
  %``The Sign Problem via Imaginary Chemical Potential,''
  Phys.\ Rev.\  D {\bf 75}, 114504 (2007)
  [arXiv:hep-lat/0703004].
  %%CITATION = PHRVA,D75,114504;%%

\bibitem{Conradi}
  S.~Conradi and M.~D'Elia,
  %``Imaginary chemical potentials and the phase of the fermionic determinant,''
  Phys.\ Rev.\  D {\bf 76}, 074501 (2007)
  [arXiv:0707.1987 [hep-lat]].
  %%CITATION = PHRVA,D76,074501;%%


\bibitem{BD}
  N. Bilic, K. Demeterfi, 
  %One-dimensional QCD with finite chemical potential,
  Phys. Lett. B, {\bf 212} (1988) 83.

\bibitem{RV}
  L.~Ravagli and J.~J.~M.~Verbaarschot,
  %``QCD in One Dimension at Nonzero Chemical Potential,''
  arXiv:0704.1111 [hep-th].
  %%CITATION = ARXIV:0704.1111;%%

\bibitem{CFZ} J.B. Conrey, D.W. Farmer and  M.R. Zirnbauer,
[arXiv:math-ph/0511024], 2005.

\bibitem{Martin-pr}M.R. Zirnbauer, private communication, January 2009.
\bibitem{GSS}
  M.~Golterman, Y.~Shamir and B.~Svetitsky,
  %``Breakdown of staggered fermions at nonzero chemical potential,''
  Phys.\ Rev.\  D {\bf 74}, 071501 (2006)
  [arXiv:hep-lat/0602026].
  %%CITATION = PHRVA,D74,071501;%%

\bibitem{splitrev}K. Splittorff
  %``Lattice simulations of QCD with mu(B) not = 0 versus phase quenched  QCD,''
  arXiv:hep-lat/0505001;
  %%CITATION = HEP-LAT/0505001;%%
  PoS {\bf LAT2006} 023, 
 %``The sign problem in the epsilon-regime of QCD,''
  arXiv:hep-lat/0610072.
  %%CITATION = HEP-LAT 0610072;%%.


\end{thebibliography}
\end{document}